\documentclass{AR/ar-1col-S2O}
\usepackage[numbers]{natbib}
\usepackage{url}
\usepackage{graphicx,epsfig}
\usepackage{amsfonts,amssymb,amsmath,amsthm}

\jname{Annu. Rev. Control Robot. Auton. Syst.}
\jvol{7}
\jyear{2023}
\doi{10.1146/annurev-control-062323-102456}

\usepackage[english]{babel}

\usepackage{amsmath}
\usepackage{graphicx}
\usepackage[colorlinks=true, allcolors=blue]{hyperref}

\def\Re{{\mathbb{R}}}
\def\Nat{{\mathbb{N}}}
\def\Z{{\mathbb{Z}}}

\def\histi{{\tilde{\iota}}}

\def\histy{{\tilde{y}}}
\def\histY{{\tilde{Y}}}
\def\histu{{\tilde{u}}}

\def\E{{\cal{E}}}

\def\hb{{\bar{h}}}
\def\fb{{\bar{f}}}
\def\Hb{{\bar{H}}}
\def\Fb{{\bar{F}}}

\def\c{\colon}
\def\ra{\rightarrow}

\def\A{{\cal{A}}}
\def\omx{{\alpha}} 
\def\is{{\iota}}  
\def\histo{{\tilde{\omega}}}
\def\histO{{\tilde{\Omega}}}
\def\histi{{\tilde{\iota}}}

\def\imap{{\kappa}}

\def\C{{\cal C}}

\def\K{{\cal K}}

\def\atan2{\operatorname{atan2}}
\def\pow{{\rm pow}}
\def\sign{{\rm sign}}

\def\PR{{\rm PR}}
\def\FFM{{\rm FFM}}

\def\I{{\cal I}} 
\def\ifs{{\cal I}} 
\def\Ifs{{\cal I}} 

\def\ifsder{{\cal I}_{der}} 
\def\ifsndet{{\cal I}_{ndet}} 
\def\ifsprob{{\cal I}_{prob}} 
\def\ifshist{{\cal I}_{hist}} 

\usepackage{acronym}
\acrodef{Ispace}[I-space]{\emph{information space}}
\acrodef{Istate}[I-state]{\emph{information state}}
\acrodef{Imap}[I-map]{\emph{information mapping}}
\acrodef{ITS}[ITS]{\emph{information transition system}}
\acrodef{ITSs}[ITSs]{\emph{information transition systems}}
\acrodef{DITS}[DITS]{\emph{deterministic information transition system}}
\acrodef{NITS}[NITS]{\emph{nondeterministic information transition system}}
\acrodef{POMDPs}[POMDPs]{\emph{partially observable Markov decision processes}}
\acrodef{PSRs}[PSRs]{\emph{predictive state representations}}

\newtheoremstyle{modelsty}
  {\topsep}
  {\topsep}
  {\rm}
  {0pt}
  {\bfseries}
  {\newline}
  {10mm}
  {}

\theoremstyle{modelsty}

\newtheorem{example}{Example}
\newcommand{\qex}{\hfill $\blacksquare$ \hfill \\}

\begin{document}

\markboth{LaValle et al.}{Perception Engineering}

\title{From Virtual Reality\\ to the Emerging Discipline of Perception Engineering}
\author{Steven M. LaValle, Evan G. Center,\\ Timo Ojala, Matti Pouke, Nicoletta Prencipe,\\ Basak Sakcak, Markku Suomalainen,\\ Kalle G. Timperi, and Vadim K. Weinstein}

\begin{abstract}
This paper makes the case that a powerful new discipline, which we term {\em perception engineering}, is steadily emerging.  It follows from a progression of ideas that involve creating illusions, from historical paintings and film, to video games and virtual reality in modern times.  Rather than creating physical artifacts such as bridges, airplanes, or computers, {\em perception engineers} create illusory perceptual experiences.  The scope is defined over any agent that interacts with the physical world, including both biological organisms (humans, animals) and engineered systems (robots, autonomous systems).  The key idea is that an agent, called a {\em producer}, alters the environment with the intent to alter the perceptual experience of another agent, called a {\em receiver}.  Most importantly, the paper introduces a precise mathematical formulation of this process, based on the von Neumann-Morgenstern notion of information, to help scope and define the discipline.  It is then applied to the cases of engineered and biological agents with discussion of its implications on existing fields such as virtual reality, robotics, and even social media.  Finally, open challenges and opportunities for involvement are identified.  

\vspace*{0.5cm}

The authors are with the Faculty of Information Technology and Electrical Engineering, University of Oulu, Finland.



\end{abstract}


\maketitle

\section{INTRODUCTION}\label{sec:intro}



The field of virtual reality (VR) focuses mainly on a raft of
technologies that when carefully assembled causes its users to have an
illusory perceptual experience.  When we refer to VR in this paper, we include augmented reality and other extended realities, which are becoming quickly unified as technologies advance.  
Most commonly, a VR user wears a head-mounted display (HMD) that provides visual and acoustic
stimulation that is adjusted to her frame of reference using sensors
that track head and other body movements.  One goal is to obtain a sense of {\em
  presence}, which combines the place illusion (feeling as if in
another location or world) and plausibility (feeling as if the virtual events were really taking place) \cite{slater2022separate}.  Although the term `virtual reality' can be traced back to the philosopher Immanuel Kant \cite{Von96}, with its current usage proposed by Jaron Lanier in the 1980s, it remains elusive to precisely define it in a way that is not wholly dependent on the engineered devices of the times.  

Regrettably, the industries surrounding VR have struggled for several decades
through hype cycles of high expectations and investment followed by
periods of disillusionment as the contemporary technologies are unable
to deliver on promises.  To bring about stable progress from a
long-term research perspective, the challenge is to understand
precisely what VR is and how it works on the entire system involved: A
combination of engineered devices (displays, sensors) and a biological
organism (human or otherwise).  Having a rigorous scientific
foundation could help improve design and analysis of VR systems so
that steady, predictable progress can be made toward achieving a
better quality of experience.  To advance in this direction, we claim
that VR, as it is viewed today, is merely one instance of a larger and
steadily emerging discipline, which we term {\em perception engineering}.  Over the coming years, we expect a rise in methods that create targeted perceptual illusions or experiences, much more broadly than the way HMD-centered VR is imagined today, and supported by fields such as machine learning, nanophotonics, and even social media technologies, in addition to the usual reliance of computer graphics, computer vision, sensors, computing systems, and displays.  As will be explained shortly, perception engineering must also be based on principles from the biological sciences.

\paragraph*{Why engineering?} 

At its core, engineering is the intentional process
of reshaping the environment to our advantage. The oldest examples are
primitive tools and weapons which date back over 1,500,000 years, whereas later examples include the
roads, bridges, aqueducts, mills, cars, airplanes, and electronics
that made large-scale civilizations flourish. Early engineering
emphasized practical applicability, often relying on trial and error
to converge towards working solutions. However, modern engineering
research has adopted the scientific method to not only construct a
working solution, but to theorize about what could be a better
solution to a particular problem or class of problems.  This process
usually involves developing mathematical models, reviewing the
state-of-the-art, generating predictive hypotheses, collecting new
data, iterating designs, and eventually sharing the findings in
peer-reviewed articles. 
We want to leverage the benefits of this well-proven methodology, but what is the engineered
artifact in our setting that would be analogous to, say, an airplane?
We argue that it is a targeted {\em perceptual experience}, and VR
devices such as HMDs are merely a component of a system that achieves
it.  We thus contemplate what it means to engineer a perceptual
experience.  Through this shift in understanding, we move the focus
more toward the organism that has an experience, and away from
particular devices that rapidly come and go.  This means that in addition to leveraging the physical sciences to engineer devices, we must also leverage the biological sciences, especially perceptual psychology, neuroscience, and physiology.

\paragraph*{Why perception?}

People have been creating perceptual illusions for millennia as well,
with the oldest known cave paintings dating back over 45,000 years.  Through imagination, trial-and-error, and
skillfully leveraging available technologies, artists continually
develop more impressive works that seem to fool our senses and
stimulate our imagination.  For example, when the perspective method
emerged in the 15th century, paintings depicted
imagined worlds with consistent perceptions of depth and scale.  Even
more impressive is that exploiting the {\em stroboscopic apparent
  motion} effect \cite{Prince2010}, while being shown a rapid sequence of
pictures, has led to over a century of cinema.  In recent decades,
simulators, video games, and VR have enabled active perceptual experiences,
in which users interact with virtual environments, rather than
passively consuming artistic content.  Thus, we wonder what applying
engineering principles to the design, analysis, and delivering of
perceptual experiences will bring, as a natural complement to the
works of artists.

\paragraph*{The value of modeling}

To embark on this journey, an important step is to develop
mathematical models that accurately capture what is understood so far,
while giving insights into what may or may not be possible to achieve
in the future.  Mathematical models enable engineers to analyze or
predict what {\em would happen} if a proposed design were constructed,
so that the engineering process is accelerated.  They also yield
characterizations of what is expected to happen for systems that are
actually built.  Furthermore, especially for robots and autonomous
systems, the models form the basis of algorithms or control laws that
are implemented on devices.  Perhaps most importantly, mathematical
models form the foundations of technical disciplines that survive for
generations, which distances them from the particular technological
artifacts of the day.  For example, a general mathematical formulation
of configuration spaces is crucial in robotics for the development of
general motion algorithms and nonlinear control laws.  Control theory
itself relies on the ability to unify a vast array of physical
settings, whether mechanical, electrical, chemical, and so on, by
mathematically characterizing them as a family of parameterized
differential equations.  Similarly, Turing machines provide a precise
mathematical characterization of algorithms and a foundation upon
which the discipline of computer science has been built.  In this
paper, we propose the first full mathematical model of perception engineering, including VR.

\paragraph*{What should the model encompass?}

At the very least, we expect a mathematical formulation of perception
engineering to model a person experiencing VR by wearing an HMD.  We
will generalize, however, to allow any organism, such as
hamsters and fruit flies \cite{naik2020animals}.  As opposed to being a
biological entity, we will even allow it to be fully engineered,
as in the case of a robot.  We use the term {\em agent} to refer to
any such organism or robot, which generally has the ability to actuate
in response to external stimulation that is sensed.  Robot agents are encompassed because they can be fully modeled and analyzed, as opposed to biological organisms, which must be reverse engineered.  Thus, whereas organisms start off as a {\em black box}
(or perhaps a {\em gray box} thanks to biological sciences), engineered systems may serve as a fully explainable {\em white box}, leading to greater clarity and understanding.

Central to our formulation are two types of agents: 1) a {\em producer}, which creates and delivers a
targeted perceptual experience by altering the environment of 2) a {\em
  receiver}.  This is the essence of perception engineering: \\

{\em An agent alters the environment with the intent to alter the
  perceptual experience of another agent.} \\
  
\begin{figure}
\centerline{\hspace*{0.0in}\psfig{figure=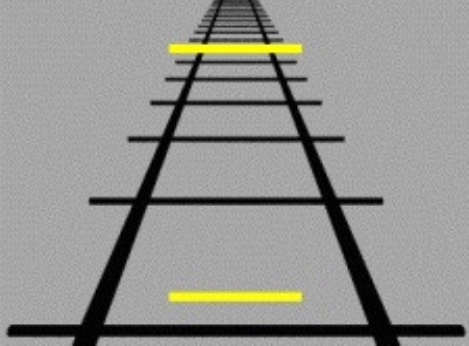,height=1.0in}\hspace*{-4.0in}\psfig{figure=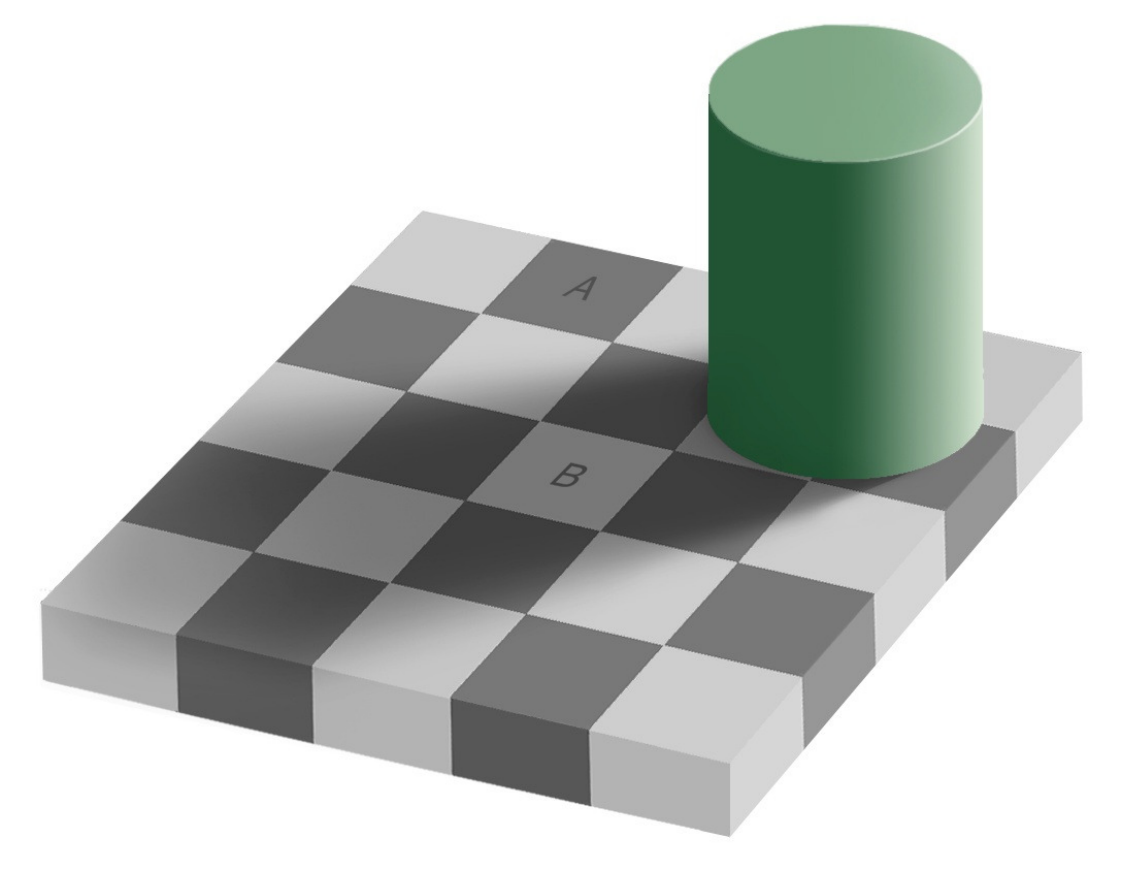,height=1.0in}\hspace*{-4.0in}\psfig{figure=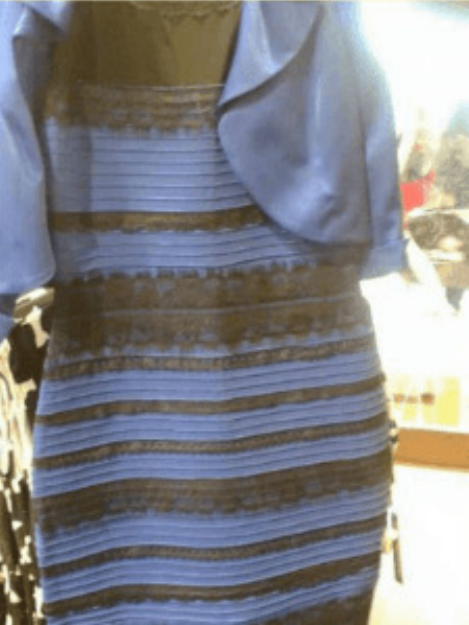,height=1.0in}\hspace*{-4.0in}\psfig{figure=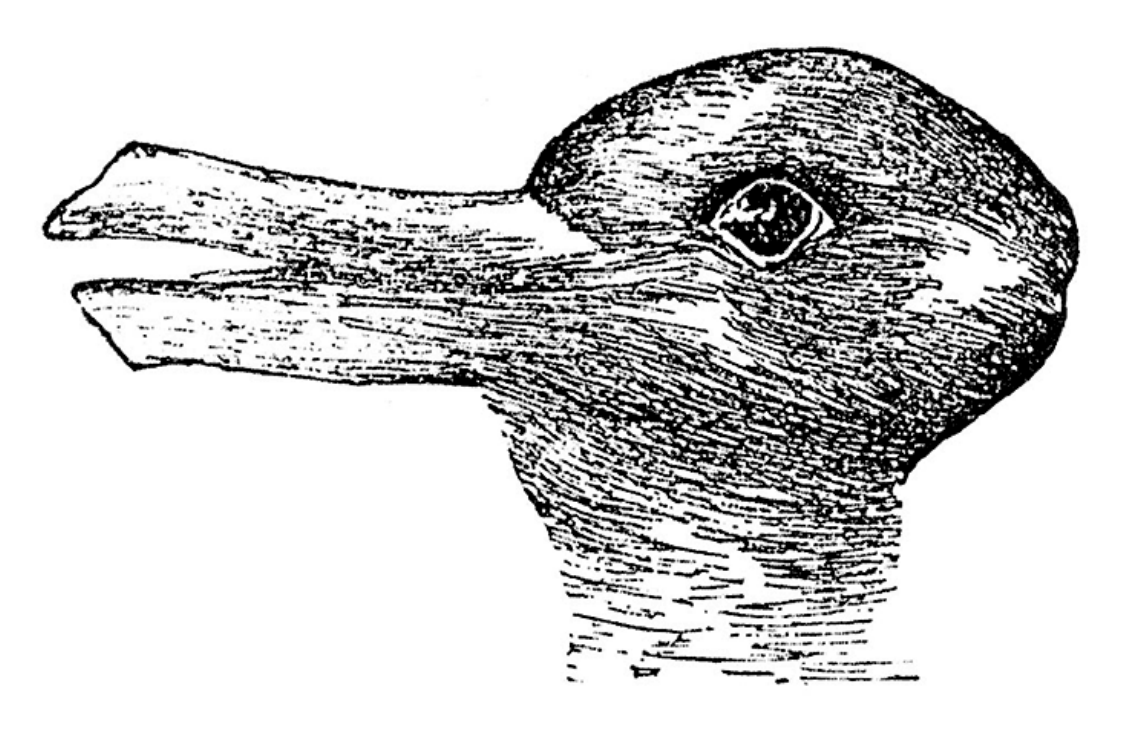,height=1.0in}}
\centerline{(a) \hspace*{1.2in} (b) \hspace*{1.0in} (c) \hspace*{1.0in} (d)}
\caption{\label{fig:ill} Several well-known illusions: (a) Ponzo (surprisingly, the yellow bars have equal length); (b) Checker Shadow (tiles A and B are surprisingly the same shade); (c) The Dress (some see it as black and blue, others as white and gold; (d) Rabbit-Duck (seems to flip between two different animals).}
\end{figure}

\noindent We also expect the targeted experience to be {\em plausible} (credible in some sense) and {\em illusory} (based on illusions), meaning that the receiver's perceptions do not match `reality'.  The notion of an illusion must be carefully defined, which is a challenging task considering how the term is used loosely.  Consider, for example, the so-called illusions in Figure \ref{fig:ill}; Section \ref{sec:illusion} will rigorously define illusions, and then apply it in Section \ref{sec:ivr} to these examples.


In addition to VR, paintings, motion pictures, and video games, we
also consider counterfeiting, magic tricks, wearing makeup, and just
plain old lying to be examples that at least nominally fall under
perception engineering.  We avoid, however, direct alteration of the
internals of the receiver, which might correspond, for example, to
brain-machine interfaces, drugs, or viruses.  We are more interested
in a perceptual experience that results purely from interactions with 
the environment via sensing and actuation.

If there are multiple producers and receivers, then perception
engineering extends naturally to a kind of social engineering.  For
example, a producer can lie to a receiver, and then the lie is `innocently'
propagated to other receivers.  We can certainly imagine that fake
news spreading on social media is a kind of virtual reality at a
societal level.  This paper focuses mainly on a single
producer and receiver, but the powerful multi-agent extension to the
social setting is also covered.

\paragraph*{The rest of this paper}

Section \ref{sec:agents} defines a general model of agents, capable of sensing and acting, in a shared environment.  Each agent is modeled as a coupled dynamical system that has internal `brain' states that interact with its external, surrounding environment states.  Section \ref{sec:core} then develops the principles by which producer agents create illusory perceptual experiences for receiver agents.  Important ideas include plausibility, robustness, forced fusion, and information-feedback policies.  Section \ref{sec:robots} applies this mathematical framework to the case of robots and other fully engineered systems.  Section \ref{sec:humans} then applies it to humans and other biological organisms, with substantial focus on VR.  Section \ref{sec:end} concludes the paper by listing the challenges and opportunities ahead for building up the field of perception engineering.


\section{AGENTS THAT SENSE AND ACT}\label{sec:agents}
 

\subsection{Defining an Agent}\label{sec:defs}



An {\em agent} refers to an entity, either biological or engineered, that has a physical body and interacts with its environment though sensing and actuation.  A clear boundary must be set between the agent's {\em internal} `brain' and the {\em external} world, which corresponds to its body and any other physical attributes that are relevant in the surrounding environment.  The {\em information space} (or {\em I-space}), $\ifs$, is defined as the (nonempty) set of all internal states; each $\is \in \ifs$ is called an {\em information state} or {\em I-state}.  
The notion of information used here is inspired by von Neumann and Morgenstern for games with imperfect information \cite{VonMor44}, and extended to robotics \cite{Lav06,SakWeiLav23}.
This is not to be confused with Shannon's notion of information, which is independent, but can be used in conjunction with our formulation.  The external {\em state space} (or {\em X-space}), $X$, is defined as the (nonempty) set of all possible physical states of the agent's body and environment.  Each $x \in X$ is called an {\em external state} or {\em X-state}.

Let $\K$ be the {\em set of stages}, which correspond to discrete time instances, with the implication that stage $k+1$ comes after stage $k$ for all $k \in \K$.  Every $k \in \K$ corresponds to some {\em stage} $k$, X-state $x_k \in X$, and I-state $\is_k \in \ifs$.  We either allow $\K$ to be infinite, in which case $\K = \Nat = \{1,2,3,\ldots\}$, or finite, in which case $\K=\{1,2,\ldots,K+1\}$ for some final stage $K+1$.

To model actuation, let $U$ be the set of {\em actions} available to the agent, through which it acts upon the external world via an {\em X-state transition function}, abbreviated as {\em XTF}, expressed as $f \c X \times U \ra X$.  An action $u_k \in U$ applied at stage $k \in \K$ from X-state $x_k \in X$ results in a transition to X-state $x_{k+1} = f(x_k,u_k)$ at stage $k+1$.  

Sensing is modeled via a {\em sensor mapping}, $h \c X \ra Y$, in which $Y$ is a set of possible sensor {\em observations}.  At each stage $k$, an observation $y_k = h(x_k)$ is produced, using the X-state $x_k$ at stage $k$.  Extensions are possible, such as being action-dependent, time-dependent, or even based on a history of states as in the case of odometry \cite{Lav06,Lav12b}.

The {\em I-state transition function}, {\em ITF}, is $\phi\colon \Ifs \times U \times Y \rightarrow \Ifs$.  To make the agent into an autonomous (fully determined) system, suppose that its action at each stage depends only on its I-state.  This is expressed as a {\em policy} $\pi \colon \Ifs \rightarrow U$, which eliminates the $U$ component from the domain of $\phi$ and results in $\phi_\pi \colon \Ifs \times Y \rightarrow \Ifs$ (because $u_k = \pi(\is_k)$ is determined from the I-state $\is_k$).  Putting the definitions together results in a coupled dynamical system:
\begin{align}\label{eqn:sys}
  \centering
  \is_k &=\phi_\pi(\is_{k-1},y_k) &\text{in which } y_k & = h(x_k), \mbox{ and}\nonumber \\
  x_{k+1} &=f(x_k, u_k)     &\text{in which } u_k & = \pi(\is_k).
\end{align}

\begin{example}[Intbot]\label{ex:intbot} Let $X = \Z$, the set of all integers.  Let $U = \{-1,0,1\}$, and the XTF is defined as $x_{k+1} = f(x_k,u_k) = x_k + u_k$.  To create a perfect-information setup, let $Y=\Ifs=\Z$, $y_k = h(x_k) = x_k$ and $\is_{k+1} = \phi_\pi(\is_k,y_{k+1}) = y_{k+1} = x_{k+1}$.  The policy $\pi \c \ifs \ra U$ is then expressible as $u_k = \pi(\is_k) = \pi(x_k)$.  A {\em stabilizing} policy is $\pi(x_k) = -\sign(x_k)$ (yielding zero if $x_k = 0$), which brings the state to zero.  From any initial X-state, $x_1$, the system will arrive at $x_k = 0$ at stage $k = |x_1|+1$, and remain there forever.  It counts down to zero.  \qex
\end{example}

\noindent The far more common and interesting case is when the sensor mapping $h$ is a many-to-one mapping.  Generally, for a given observation $y_k \in Y$, the {\em preimage}
\begin{equation}\label{eqn:preimage}
h^{-1}(y_k) = \{x_k \in X \;|\; y_k = h(x_k)\}
\end{equation}
indicates the set of all possible X-states that could have caused it.  Many-to-one ambiguity usually forces the X-state and I-state to have a non-trivial correspondence, to be explained in Section \ref{sec:ispaces}.

The models so far assume that the next state $x_{k+1}$ is completely predictable from $x_k$ and $u_k$, and the sensor observation $y_k$ is completely predictable from $x_k$.  We sometimes want to remove this limitation by defining a {\em disturbance-based} model.  There will be two possibilities.  The first is {\em nondeterministic disturbance}, in which case $f$ and $h$ are replaced by functions that produce a {\em set} of possible outcomes, rather than a single outcome.  The {\em nondeterministic XTF} is $F \c X \times U \ra \pow(X)$, in which $\pow$ denotes the power set, and $F$ is the replacement of $f$.  Thus, for a given $x_k$ and $u_k$, $F(x_k,u_k) \subseteq X$ yields a nonempty set of possible $x_{k+1}$.  The {\em nondeterministic sensor mapping} is $H \c X \ra \pow(Y)$, and replaces $h$.  Given $x_k$, $H(x_k) \subseteq Y$ yields a nonempty set of possible observations $y_{k+1}$.  The other possibility is {\em probabilistic disturbance}, in which case $f$ and $h$ are replaced by functions that each produce a {\em probability density function (pdf)} (under appropriate measurability assumptions) on the space of possible outcomes.  The {\em probabilistic XTF} is the pdf $p(x_{k+1}\;|\;x_k, u_k)$, and the {\em probabilistic sensor mapping} is the pdf $p(y_k\;|\;x_k)$.  The general possibilities for $\ifs$ and $\phi$, and including extensions to nondeterministic and probabilistic disturbances are presented after the following example.

\begin{example}[Linebot]\label{ex:linebot}
We extend Example \ref{ex:intbot} to $X = \Re$, the real number line.  Let $U = \{-1,0,1\}$ and $x_{k+1} = f(x_k,u_k,\theta_k) = x_k + u_k + \theta_k$, which includes some disturbance parameter $\theta_k \in [-1/2,1/2]$.  If $\theta_k$ is modeled nondeterministically, then $F(x_k,u_k) = [x_k+u_k-1/2,x_k+u_k+1/2]$ produces an interval of next possible states.  If it is modeled probabilistically, then suppose a pdf $p_\theta(\theta_k)$ is given over the interval $[-1/2,1/2]$;  $p(x_{k+1}\;|\;x_k,u_k)$ can then be defined as $p_\theta(x_{k+1} - x_k - u_k)$.  As an example of adding disturbance to the sensing model, $y_k = h(x_k,\psi_k) = x_k + \psi_k$, which includes some disturbance parameter $\psi_k \in [-1,1]$.  The nondeterministic sensor model would be $H(x_k) = [x_k-1,x_k+1]$.  The probabilistic sensor model would involve a pdf $p_\psi(\psi_k)$, and $p(y_k\;|\;x_k) = p_\psi(y_k-x_k)$.
\qex
\end{example}


\subsection{Internal Information State Transitions}\label{sec:ispaces}


The coupled system (\ref{eqn:sys}) allows almost anything for the agent's internal system, I-space $\ifs$ and ITF.  What could they be?  First consider two extreme possibilities.  If $h$ is defined as $y = h(x) = x$ with $Y = X$, then the X-state is perfectly sensed at every stage.  We could then write $\ifs = X$ and $\phi$ simply mirrors $f$.  This corresponds to an agent that conditions its actions on the precise X-state.  Thus, its policy, called {\em state-feedback}, takes the form $\pi\colon X \rightarrow U$.  At the other extreme, we could make $\ifs = \{0\}$, a singleton that is completely insensitive to X-state variations.  Only a {\em constant} policy is possible: $\pi(0) = u$ for some particular action $u \in U$.

The interesting cases lie between these extremes and address the crucial question:  To function appropriately, what should an agent retain as I-states?  It is convenient to refer to {\em memory}, which merely means that the I-state depends on at least some information regarding actions and observations from prior stages.  The singleton I-space is clearly memoryless, but a more interesting case is to make $\ifs = Y$ and let $\phi(\is_{k-1},y_k) = y_k$.  This results in pure {\em sensor feedback}, with policies $\pi\colon Y \rightarrow U$.  To add a small amount of memory, let $\ifs = \K$ and $\phi(\is_{k-1},y_k) = k$, resulting in {\em stage feedback} policies $\pi \colon \K \rightarrow U$ (alternatively known as {\em open-loop}).    The stage-feedback case at least uses the information of how many stages have passed.

Although it may seem that any ITF, and corresponding I-space, are possible, it turns out that one important condition, known as {\em sufficiency}, must be satisfied.  It is briefly described here; see \cite{Lav06,SakWeiLav23} for more.  Consider all information that might be available to an agent after $k$ stages.  Let $\eta_k$ be called the {\em history I-state}, defined as $\eta_k = (\histu_{k-1},\histy_k)$ in which $\histu_{k-1} = (u_1,u_2,\ldots,u_{k-1})$ and $\histy_k = (y_1,y_2,\ldots,y_k)$.  Let $\eta_1 = y_1$.  The set of all history I-states is itself an I-space, denoted by $\ifshist$.  This corresponds to perfect, complete memory, with policies taking the form $\pi\colon\ifshist\rightarrow U$.  Now imagine trying to collapse or compress the history I-states to create a {\em derived I-space} $\ifsder$ via an {\em information mapping} $\imap\colon\ifshist\rightarrow\ifsder$ \cite{Lav06}.  All I-spaces discussed so far can be obtained in this way.  For a well-defined ITF $\phi_{der}$ to exist, the sufficiency condition is that
$\imap(\phi_{hist}(\imap^{-1}(\is_k),u_k,y_{k+1}))$ is a singleton.  In other words, $\phi_{der}$ can be defined so that a unique I-state is obtained and is consistent with what would have been calculated from retaining full histories.  It was shown in \cite{WeiSakLav22} that minimal sufficient information mappings, and corresponding ITFs, exist and are unique in very general settings.

We now present two alternative {\em model-based} families of sufficient ITFs, called {\em nondeterministic} and {\em probabilistic}.  They are considered model-based because the ITF is expressed in terms of $X$, $f$, and/or $h$.   Of the ITFs presented so far, only state feedback has been model-based and is a special case of the nondeterministic family.  The I-space is $\ifsndet = \pow(X)$.  The ITF $\phi_{ndet}$ incrementally calculates the set of possible X-states at each stage.  For each $k \in \K$, let $X_k(\eta_k)$ denote the set of possible X-states at stage $k$ given the history I-state $\eta_k$.  Suppose at the first stage, $X_1(\eta_1) = h^{-1}(y_1)$ (using the preimage from (\ref{eqn:preimage})).  The ITF calculates $X_{k+1}(\eta_{k+1})$ using only $X_k(\eta_k)$, $u_k$, and $y_{k+1}$; thus, it takes the form
\begin{equation}
X_{k+1}(\eta_{k+1}) = \phi_{ndet}(X_k(\eta_k),u_k,y_{k+1}) .
\end{equation}
Assuming inductively that $X_k(\eta_k)$ is given, consider the possible X-states at stage $k+1$ under the application of action $u_k$:
\begin{equation}\label{eqn:ndet1}
X_{k+1}(\eta_k,u_k) = \{ x_{k+1} \in X \mid \mbox{ there exists } x_k \in X_k(\eta_k)\mbox{ for which } x_{k+1} = f(x_k,u_k) \} .
\end{equation}
Once the new sensor observation $y_{k+1}$ arrives, the preimage $h^{-1}$ is used to constrain the set of possible states, resulting in:
\begin{equation}\label{eqn:ndet2}
X_{k+1}(\eta_{k+1}) = X_{k+1}(\eta_k,u_k,y_{k+1}) =  X_{k+1}(\eta_k,u_k) \cap h^{-1}(y_{k+1}) .
\end{equation}
This completes the definition of $\phi_{ndet}$.  Note that state feedback is a special case in which $X_k(\eta_k)$ is a singleton for all $k \in \K$.  We can easily extend $\phi_{ndet}$ to account for nondeterministic disturbances by replacing $x_{k+1} = f(x_k,u_k)$ by $x_{k+1} \in F(x_k,u_k)$ in (\ref{eqn:ndet1}) and $h^{-1}$ by $H^{-1}$ in (\ref{eqn:ndet2}), in which $H^{-1} = \{ x_{k+1} \in X \mid y_{k+1} \in H(x_{k+1}) \}$.  Note that $f$ and $h$ (or $F$ and $H$) are used internally by the agent, and they might be inconsistent with the actual external world; this will be clarified in Section \ref{sec:multi} and is critical for defining illusions.

We now define the probabilistic model-based family.  The corresponding I-space is $\ifsprob$, which is the set of all probability density functions (pdfs) over $X$ (again, under appropriate measurability assumptions).  Probabilistic disturbance-based replacements of $f$ and $h$ are given as $p(x_{k+1}|x_k,u_k)$ and $p(y_k|x_k)$, respectively.  Let $p(x_k|\eta_k)$ denote the pdf of the state at stage $k$ given $\eta_k$ (the probabilistic counterpart to $X_k(\eta_k)$).  The ITF takes the form
\begin{equation}
p(x_{k+1}|\eta_{k+1}) = \phi_{prob}(p(x_k|\eta_k),u_k,y_{k+1}) ,
\end{equation}
and is equivalent to Bayesian filtering and the basis of POMDPs.  The process is started by a given prior $p(x_1)$.  The probabilistic counterpart to (\ref{eqn:ndet1}) is marginalization, resulting in:
\begin{equation}
p(x_{k+1} \mid \eta_k, u_k) = \int_X p(x_{k+1} \mid x_k, u_k) \; p(x_k \mid \eta_k) \; dx_k .
\end{equation}
The probabilistic counterpart to the intersection in (\ref{eqn:ndet2}) is Bayes' rule, resulting in
\begin{equation}
p(x_{k+1} \mid \eta_{k+1}) = p(x_{k+1} \mid \eta_k, u_k, y_{k+1}) = \frac{
p(y_{k+1} \mid x_{k+1}) \; p(x_{k+1} \mid \eta_k, u_k) }{\int_X p(y_{k+1} \mid x_{k+1}) \; p(x_{k+1} \mid \eta_k, u_k) \; dx_{k+1}}.
\end{equation}
Conditional independence assumptions and further details are explained in \cite{Lav06}.  The celebrated Kalman filter is a special case (linear systems with Gaussian disturbances) in which the I-states become trapped in a low-dimensional subspace of $\ifsprob$, and the ITF can be calculated using matrix algebra.  Each I-state then corresponds to mean and covariance of the X-state at stage $k$.

\begin{figure}
\centerline{\psfig{figure=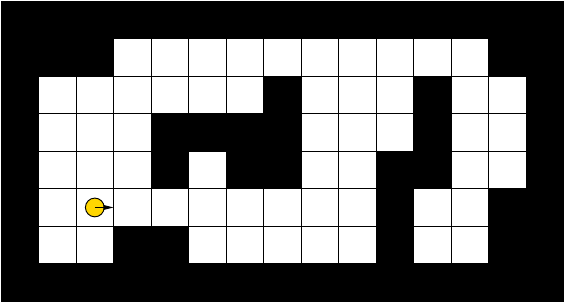,width=2.0in}\hspace*{-2.5in}\psfig{figure=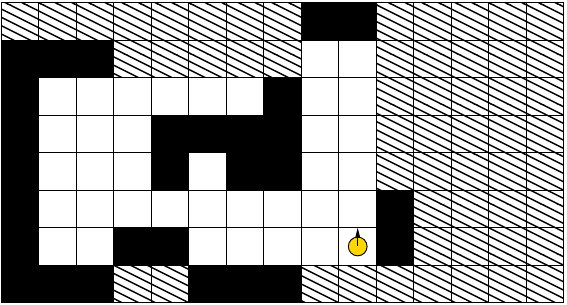,width=2.0in}}
\centerline{(a) \hspace*{2.5in} (b)}
\caption{\label{fig:bkgrid} (a) A discrete grid problem is made in
which a robot is placed into a bounded, unknown environment.  (b) An
encoding of a partial map, obtained from some exploration.  The
hatched lines represent unknown tiles (neither white nor black).}
\end{figure}

\begin{example}[Gridbot]\label{ex:gridbot}
A mobile robot moves on a 2D grid and can face in one of four orientations (up, down, left, right), as shown in Figure \ref{fig:bkgrid}(a).  At each possible $(i,j) \in \Z \times \Z$ position there is a {\em tile}, which may be either {\em black} or {\em white}.  If it is white, then the robot can occupy its position; otherwise, it is blocked.  The robot starts on one tile among a finite, unknown, connected set of white tiles.  All other tiles are black.  Each possible set of white tiles is called an {\em environment}.  The X-space is $X \subseteq \Z^2 \times D \times \E$, in which $D = \{0,1,2,3\}$ is the set of four directions and $\E$ is the set of all possible environments.  An X-state $x \in X$ can be expressed as $(i,j,d,E)$ in which $(i,j) \in E$, $d \in D$, and $E \in \E$. There are two actions $U=\{0,1\}$, in which $u=0$ causes the robot to rotate 90 degrees counterclockwise, and $u=1$ makes it attempt to move forward one tile in the direction it is facing (it does not move if blocked by a black tile).  A `depth' sensor $y_k = h(x_k)$ reports the distance, in terms of number of white tiles, to first black tile in the direction the robot faces; thus, $Y = \{0,1,2,\ldots\}$.  Using (\ref{eqn:ndet1}) and (\ref{eqn:ndet2}), the set $X_k(\eta_k)$ of possible X-states is calculated after each action and observation, respectively; however, it is important to encode $X_k(\eta_k)$ compactly, rather than list all possible states for an infinite collection of environments.  Initially, $X_1(\eta_1)= h^{-1}(y_1)$.  During exploration, tiles sensed to be white or black are recorded using $(i,j)$ coordinates, with $(0,0)$ as the initial white tile.  All others may be labeled as gray, meaning unknown or unexplored; see Figure \ref{fig:bkgrid}(b).  This is a highly compressed encoding of $X_k(\eta_k)$, which technically belongs to a derived I-space of such encodings.  To obtain a nondeterministic disturbance model, $h$ may produce a set of possible distances, and the calculations in (\ref{eqn:ndet2}) use the larger preimages $H^{-1}(y_{k+1})$.  A Bayesian version can be made by introducing probabilistic alternatives to $f$ and $h$, and using $\phi_{prob}$ to calculate probabilistic I-states; see \cite{ThrBurFox05} for details.  \qex
\end{example}

\subsection{Defining Plausibility and Illusions}\label{sec:illusion}

Here we address the possibility that the models used in the ITF do not perfectly coincide with `reality' as the agent interacts with its environment.  Of course, reality itself will be defined as a model, but it is nevertheless crucial to maintain a distinction.  Thus, we refer to $X$, $f$, and $h$ used in the agent's ITF as {\em intrinsic} models.  To provide an outside frame of reference, we will introduce their counterparts $\Omega$, $\fb$, and $\hb$, and refer them as {\em extrinsic} models.  If there is no disagreement between the intrinsic and extrinsic models, then $\Omega = X$, $\fb = f$, and $\hb = h$; this distinction was not yet needed in Section \ref{sec:ispaces}.  Discrepancies between the intrinsic and extrinsic models will be crucial for modeling perceptual illusions.

Let $\Omega$ be called the {\em universe space}, or just the {\em universe}, which models the set of all possible physical states of the world from a third-person or god-like perspective.  At each stage $k \in \K$, the universe state is $\omega_k \in \Omega$ and the {\em universe transition function}, {\em UTF}, is defined as $\fb \c \Omega \times U \ra \Omega$. Similarly, the {\em universe sensor mapping} is defined as $\hb \c \Omega \ra Y$.



Next, we define a general way to model potential correspondences between X-states and universe states from a third-person perspective (outside of the agent).  For any $X$ and $\Omega$, let $C \subset X \times \Omega$ be called a {\em correspondence relation}.  If $(x_k,\omega_k) \in C$ it is said that $x_k$ {\em corresponds to} $\omega_k$.  Note that $C$ allows for the correspondence to be one-to-one, many-to-one, one-to-many, or many-to-many.  The relation $C$ is called {\em onto} if for all $\omega_k \in \Omega$, there exists an $x_k \in X$ such that $(x_k,\omega_k) \in C$. If $C$ is one-to-many and onto, then there exists a function $\alpha \c \Omega \ra X$ such that $(\alpha(\omega_k),\omega_k) \in C$ for all $\omega_k \in \Omega$; thus, the X-state can be derived from the universe state as $x_k = \alpha(\omega_k)$.


The relationship between the agent's I-state and `reality' in the universe can be established through their relationships to $X$.  Let the {\em model relation} $M \subseteq \I \times X$ associate I-states to possible X-states.  If the nondeterministic ITF family is used, then $(X_k(\eta_k),x_k) \in M$ if and only if $x_k \in X_k(\eta_k)$.  (For probabilistic ITFs, $M$ may be defined using thresholds on pdfs to obtain probabilistic correspondences.)  An I-state $\is_k$ is called {\em implausible} if there does not exist any $x_k \in X$ such that $(\is_k,x_k) \in M$.  Thus, $X_k(\eta_k) = \emptyset$ would be an implausible I-state for nondeterministic ITFs.  A pair $(\is_k,x_k)$ or I-state $\is_k$ is called {\em plausible} if it is not implausible.  Its usage here is inspired by concepts of plausibility in VR research \cite{skarbez2017survey,slater2022separate}, but also differs in precise meaning.

By composing the model and correspondence relations, the {\em reality relation} $R \subseteq \I \times \Omega$ is defined as $(\is_k,\omega_k) \in R$ if and only if there exists $x_k \in X$ such that $(\is_k,x_k) \in M$ and $(x_k,\omega_k) \in C$.  A pair $(\is_k,\omega_k)$ is called an {\em illusion} if $\is_k$ is plausible and $(\is_k,\omega_k) \not \in R$.  Thus, the key idea of an illusion is that the agent perceives something as plausible but it does not correspond to reality.

\subsection{Multiple Agents}\label{sec:multi}

Suppose there are $n$ agents in a common environment. The $i$th agent is modeled using the components from Section \ref{sec:defs} and denoted as an 8-tuple $\A^i = (X^i,\ifs^i,U^i,Y^i,f^i,h^i,\phi^i,\pi^i)$.  Here, $X^i$, $f^i$, and $h^i$ are intrinsic, a distinction that was unnecessary in Section \ref{sec:defs}; thus, as in Section \ref{sec:illusion}, we seek their extrinsic counterparts.  The universe sensor mapping for each agent is $\hb^i \c \Omega \ra Y^i$.  The UTF must take into account the interactions between agents.  It is therefore specified centrally as $\fb \colon \Omega \times U^1 \times \cdots \times U^n \rightarrow \Omega$, as is common in dynamic game theory. If the agents' actions do not interfere with each other, then $\fb$ may be decomposable into individual $\fb^i$ functions over subspaces of $\Omega$, but this will not be assumed.  Correspondence, model, and reality relations can be defined over $X^i \times \Omega$, $\I^i \times X^i$, and $\I^i \times \Omega$, respectively.

\begin{example}[Two Gridbots]\label{ex:gridbot2}
We extend Example \ref{ex:gridbot} by placing two gridbot agents, $\A^1$ and $\A^2$, in a universe defined as $\Omega \subset \Z^4 \times D^2 \times \E$, which is the set of all $(i^1,j^1,i^2,j^2,d^1,d^2,E) \in \Omega$ that satisfy $(i^1,j^1) \in E$, $(i^2,j^2) \in E$, $E \in \E$, $d^1 \in D$, $d^2 \in D$, and $(i^1,j^1) \ne (i^2,j^2)$. Each agent has an X-space $X^i$ as defined in Example \ref{ex:gridbot} for a single robot.  Correspondence relations $C^i$ are defined for $i=1,2$. They are subsets of $X^i \times \Omega$ and are one-to-many and onto, implying the existence of functions $\alpha^i \c \Omega \ra X^i$ that are consistent with $C^i$.  Each $\alpha^i$ maps $\omega$ to $E$ and the position and direction of the $i$th agent in $E$, all of which are expressed using the local coordinates of $\A^i$ because $\A^1$ and $\A^2$ assign position $(0,0)$ and direction $0$ differently.

Each universe sensor mapping $\hb^i$ differs from $h^i$ by taking into account the other robot.  Let $\hb^i$ return the directional distance to the nearest black tile or other robot, if it is closer (the other robot is like a movable black tile).  The UTF $\fb$ can be derived to coincide with the individual $f^1$ and $f^2$ XTFs, but must specify what happens when robots attempt to move into each other.  Suppose each cannot move onto a tile occupied by the other, and the first agent has priority if they attempt to move to the same white tile at the same time.
\qex
\end{example}

For multiple agents, the I-spaces and ITFs are more challenging to model due to the effects of their interactions, whether accidental or intentional.  For Example \ref{ex:gridbot2}, what happens when one robot is blocked by the other?  It seems incorrect to label the tile as black.  Perhaps later it will discover that the tile is white.  Does its model allow $E$ to change?  Does it `know' there is another robot?  To define each ITF, an agent's intrinsic model must carefully specify what information regarding other agents is available.  Using all sources of potential information, an agent's ITF could be expressed as 
$\is^i_{k+1} = \phi^i(\is^1_k,\ldots,\is^n_k,u^1_k,\ldots,u^n_k,y^1_{k+1},\ldots,y^n_{k+1})$, in which the definition of $\phi^i$ may reference any component of the 8-tuple $\A^j$ for any agent, any $\hb^j$, any correspondence, model, and reality relations, and $\fb$.  

Consider designing one agent, say $\A^1$, as {\em omniscient}, meaning that it has access to as much information as possible.  Let $X^1 = \Omega$ and $y^1_k = h^1(x^1_k) = h^1(\omega_k) = \omega_k$, thus observing the universe state at every stage.  Its own I-state $\is^1_k$ records the histories of all actions, observations, and I-states for itself and all other agents.  This is a multiagent extension of $\ifshist$ from Section~\ref{sec:ispaces}.  Using the notion of sufficient information mappings $\kappa$ from Section \ref{sec:ispaces}, this I-space can be collapsed to smaller I-spaces as appropriate for accomplishing particular tasks.  In general, the model components of each $\A^i$, each $\hb^i$, the relations, and $\fb$ may be used to define $\phi_{der}^1$ over the derived I-space.

The information gathering capabilities of an omniscient agent may seem to be too much.  Most `ordinary' agents will have far less capabilities, but it is nevertheless important to define the extreme case as a starting point.  How could I-states of other agents be obtained?  One way is to predict them through simulation or computation using the other data and models.  If they are measured directly, then a sensor model should be formulated that includes I-states as part of the physical universe.  How could observations or actions of other agents be obtained?  Similarly, they could be estimated from other information, such as using $\fb$ and the measured $\omega_k$ and $\omega_{k+1}$ to determine the actions, or $\hb^i$ and the measured $\omega_k$ to determine the observation.  Otherwise, they should be directly sensed in an expanded universe.  Generally, each agent obtains a history I-state $\eta^i_k$, resulting in a history I-space $\ifshist^i$, which can be reduced using sufficient information mappings.  To allow precise models of observing I-states, actions, and observations, the universe should be expanded to include them; however, this technical level is beyond the scope of the current paper.

In terms of actuation, at one extreme, one agent could be {\em omnipotent}, enabling it to set any $\omega_k \in \Omega$ at any stage $k \in \K$.  We do not allow multiple omnipotent agents because their actions would be in conflict. We also do not allow it to directly set I-states, observations, or actions of other agents.  At the other extreme, it could be a passive {\em observer}.  Consider what happens when we, the modelers, try to mathematically analyze the entire system.  In this case, we are acting as an omniscient agent that is merely an observer so as not to interfere with its operation while analyzing what should happen theoretically.  All other agents are in between being omnipotent and an observer, and they may or may not be omniscient.


Extensions can also be made to account for disturbances.  In the nondeterministic case, the universe sensor mapping of the $i$th agent becomes $\Hb^i \c \Omega \ra \pow(Y^i)$.  The UTF becomes $\Fb \c \Omega \times U^1 \times \cdots \times U^n \rightarrow \pow(\Omega)$.  Similarly, for the probabilistic case, the corresponding corresponding universe sensing model and transition function are $p(y^i_k\;|\;\omega_k)$ and $p(\omega_{k+1}\;|\;\omega_k,u^1_k,\ldots,u^n_k)$, respectively.

\section{CREATING TARGETED PERCEPTUAL EXPERIENCES}\label{sec:core}

\subsection{Producers and Receivers}\label{sec:tpe}



We now adapt the multiagent model of Section \ref{sec:multi} to the case of a special agent $\A^p$ called a {\em producer} that delivers a targeted perceptual experience to another agent $\A^r$ called a {\em receiver}.  Key concepts running throughout Section \ref{sec:core} are plausibility and illusions, as defined in Section \ref{sec:illusion}.  To allow it to `fool' the receiver in some sense, the producer usually has access to more information than the receiver.  For example, the producer may have access to models $\fb$ and $\hb^r$, whereas the receiver only has $f^r$ and $h^r$.  Section \ref{sec:instant} starts with the simplest case, in which stage dependencies and transitions are suppressed: A fixed percept (receiver I-state) is created and maintained.  Section \ref{sec:op} will then extend the concepts to cover an omniscient producer that operates over multiple stages, resulting in a targeted, interactive, perceptual experience for the receiver.  Section \ref{sec:imp} will then strip the producer of its omniscience, resulting in incomplete or imperfect models of the receiver and even the producer itself.  This is more like the situation of VR applied to biological organisms, but also relevant for engineered systems.

\subsection{Producing Stationary Percepts and Illusions}\label{sec:instant}


This section temporarily drops the notion of stages to provide a useful and illustrative setup before developing more complicated scenarios.  The receiver's sensor model is $h^r \c X^r \ra Y^r$.  The nondeterministic I-space $\ifsndet$ is the set of all preimages of $h^r$.  Thus, the I-state, called a {\em percept}, is simply fixed as the preimage $\is^r = (h^r)^{-1}(y^r) \subseteq X^r$.  The producer is omniscient, and $X^p = \Omega$.  Let $(x^p,\omega) \in C^p$ if and only if $x^p=\omega$ (they are in perfect one-to-one correspondence).  The producer is able to set $\omega$ according to some given function $\fb \c U^p \ra \Omega$ (a simplified version of $\fb$ from Section \ref{sec:multi} to handle the stationary case), after which time $\omega$ remains constant.  It also has access to $\hb^r \c \Omega \ra Y^r$ and the receiver's correspondence relations $C^r$ and $M^r$.  Suppose that $C^r$ is one-to-many and onto so that $\alpha^r \c \Omega \ra X^r$ exists.  The model relation is defined as $(\is^r,x^r) \in M^r$ if any only if $x^r \in \is^r$.  This is equivalent to $\ifsndet$ restricted to a single stage in which only one observation is available.

If $\hb^r = h^r \circ \alpha^r$, then $(\is^r,\omega) \in R^r$ for all $\omega \in \Omega$. To see why, suppose the producer sets $\omega = \fb(u^p)$ for some $u^p \in U^p$, thereby causing the receiver to observe $y^r = \hb^r(\omega) = h^r(\alpha^r(\omega))$, and obtain a targeted I-state $\is^r = (h^r)^{-1}(y^r)$.  For any possible $u^p$, the resulting pair $(\is^r,x^r)$ would belong to $M^r$, and hence $\is^r$ is always plausible.  Furthermore, $(\is^r,\omega) \in R$, implying the pair is not an illusion.

The potential to create an illusion arises if $\hb^r \ne h^r \circ \alpha^r$.  This would allow some $\omega$ to produce $y^r_1 = \hb^r(\omega)$ and $y^r_2 = h^r(\alpha^r(\omega))$ such that $y^r_1 \ne y^r_2$.  The preimages $\is_1 = (h^r)^{-1}(y^r_1)$ and $\is_2 = (h^r)^{-1}(y^r_2)$ are distinct and disjoint.  The corresponding X-state would be $x^r = \alpha^r(\omega)$, with $x^r \not \in \is_1$ and $x^r \in \is_2$.  Thus, if $\is_2^r \ne \emptyset$ and $(\is_2^r,\omega) \not \in R$, then the pair is an illusion as defined in Section \ref{sec:illusion}.

\begin{example}[Moving a Landmark]\label{ex:1d}
Let $X^r = \Z$ and $\Omega = X^p = X^r \times \Z$, as an omniscient producer.  The correspondence relation $C^r$ is one-to-many and onto, and $\alpha^r(\omega) = x_1$, denoting $\omega = (x_1,x_2)$.  The receiver's sensor model is $y^r = h^r(x) = x^r$, which measures its state perfectly.  The producer is given the extrinsic sensor mapping $\hb^r \colon \Omega \rightarrow Y^r$, defined as $y^r = \hb^r(x_1,x_2)=x_2 - x_1$, which is the signed distance from $x_1=x^r$ to the location $x_2$ of a movable reference point or `landmark'.  The producer can thus create an illusion for the receiver that its position is any $x^r \in X^r$ by changing $x_2$.  From a third-person perspective, we could interpret the receiver's sensor model as reporting the signed distance to a tower fixed at $0 \in X^r$, but $x_2$ corresponds to the true position of the tower.  In this case, the producer moves the tower position, which is beyond the receiver's model.  Its I-state $\is^r = (h^r)^{-1}(y^r) = \{x'\}$ therefore implies that its perceived own position $x' \in X^r$ is wrong, which is an illusion.  Note that $\hb^r(\omega) \ne h^r(\alpha^r(\omega))$, as required. \qex
\end{example}

\begin{figure}
\centerline{\hspace*{-0.5in}\psfig{figure=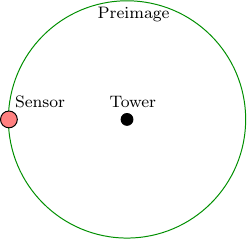,width=1.0in}\hspace*{-3.2in}\psfig{figure=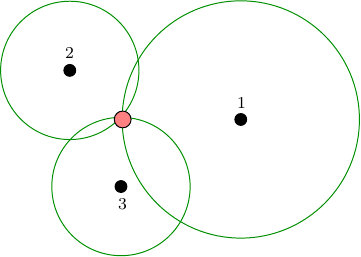,width=1.6in}\hspace*{-3.0in}\psfig{figure=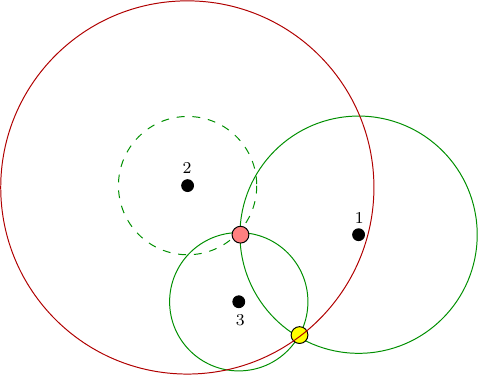,width=1.6in}}
\centerline{(a) \hspace*{1.6in} (b) \hspace*{2.0in} (c)}
\caption{\label{fig:tri} (a) A receiver's sensor measures the distance to a tower, resulting in circular preimages as I-states.  (b) For three towers, the correct position is given by the intersection of three preimages.  (c)  A producer can change the signal intensity of the second tower to create an illusory perceived position for the receiver.
}
\end{figure}

\begin{example}[Trilateration Tricks]\label{ex:tri}
Planar localization is performed by a receiver, in which $X^r = \Re^2$, by the principle of {\em trilateration}. The receiver's sensor observation $y^r$ reports the distances to one or more `towers' that serve as known landmarks with fixed position in $\Re^2$.  Using the reported distance to a single tower, the I-state $\is^r=(h^r)^{-1}(y^r)$ narrows the position down to a circle (see Figure \ref{fig:tri}(a)) of radius $y^r$.  If there are $n$ towers, then $y^r = (d^o_1,\ldots,d^o_n)$ is a vector of oberved distances to each tower.  Figure \ref{fig:tri}(b) shows the case of three towers, in which the I-state results in a unique receiver position by intersecting the three circular preimages, one for each distance measurement.


Now enters the producer, which exploits an extrinsic sensor model $y^r=\hb^r(\omega)$. The universe state is $\omega=(x_1,x_2,r_1,\ldots,r_n) \in \Omega$, in which $(x_1,x_2)$ is the receiver X-state, $r_i$ is the transmitted radio signal intensity of $i$th tower, and $\Omega \subset \Re^{n+2}$. 
The observation $d^o_i$ is actually based on the inverse-square law, $r^o_i=r_i/d^2_i$, in which $d_i$ is the actual distance to the $i$th tower and $r^o_i$ is the measured intensity at the receiver position.
Let $r^c=(r^c_1, \dots, r^c_n)$ be the vector of intensities that the sensor is calibrated to. Each element of $y^r$ is then obtained as
$d^o_i=\sqrt{r^c_i/r^o_i}$ if $r^c_i >r^o_i$ otherwise, $d^o_i=\#$, which indicates that the signal intensity is not within the interval of observable values.
Note that observing $r^0_i=r^c_i$ implies that the receiver is on the $i$th tower, which is not allowed. To yield any desired distance measurement $d'_i$ so that $d^o_i$ becomes $d'_i$, the producer changes the transmitter power $r_i$ from $r^c_i$ to $r'_i = r^c_i (d_i / d'_i)^2$. For example, if $r^c_i = 1$, then $r^o_i = 1/4$ at distance $d_i = 2$.  Changing the transmitted intensity to $r'_i = 2$ would cause the distance observation to be $d^o_i = \sqrt{2}$ ($r^o_i=1/2$), which is incorrect. Setting $r'_i \geq 4$ would then produce implausible I-states because $r^0_i\geq r_i=1$, and the respective circle will be undefined.


The producer X-space is $X^p=\Omega$.
The function $\alpha^r$ converts the producer X-state to the receiver X-state, and $C^r$ is defined accordingly.  The producer action space is $U^p = (0,\infty)^n$, and is used in $f$ to set each radio intensity $r_i$ to any positive value, inducing a desired $d'_i$ so that $y^r=(d'_1,\ldots,d'_n)$.  In the case of $n=1$, the producer interference can create receiver I-states that are circles of any radius or distance from the single tower.  For $n=2$, there is a bounded range of radio intensities (values of $u^p$) for which the circular preimages intersect; if they are disjoint, then the illusion becomes implausible (there is no receiver position that would account for $y^r$, and $\is^r = \emptyset$).  For $n=3$, shown in Figure \ref{fig:tri}(c), any small perturbation of $u^p$ results in implausibility, $\is^r = \emptyset$. \qex
\end{example}


From Example \ref{ex:tri} two kinds of concepts become clear:  1) If the I-state is plausible, then how much can the observation be perturbed while maintaining plausibility? 2) If the I-state is not plausible, then how little can the observation be perturbed to make it plausible?  Both of these concepts rely on a notion of distance in $Y^r$.  Thus, assume $Y^r$ is a metric space with metric $\rho_{Y^r} \c Y^r \times Y^r \ra [0,\infty)$.  Let $B_{Y^r}(y,d) \subseteq Y^r$ denote an open ball of radius $d$, centered at $y$. Thus, $B_{Y^r}(y,d) = \{y' \in Y^r \;|\; \rho_{Y^r}(y,y') < d \}$.

The first concept is defined as follows.  Let $P \subseteq Y_r$ be the set of all observations that yield plausible I-states, defined using the model relation $M^r$.  The {\em plausibility robustness}, $\PR_{Y^r}(y^r)$, is the largest $d$ such that $B_{Y^r}(y^r,d) \subseteq P$.  The second concept is {\em forced-fusion magnitude}, $\FFM_{Y^r}(y^r)$, which is the largest $d$ such that $B_{Y^r}(y^r,d) \subseteq Y^r \setminus P$.  In other words, it is the distance to the nearest receiver observation that would yield plausibility.  The term is inspired by its use to account for adverse symptoms in VR usage \cite{HilErnBanLan02}.  


We would like to express these concepts from the producer's perspective.  For plausibility robustness, how much can the producer vary $u^p$ and still maintain plausibility?  For forced-fusion magnitude, how much does the producer need to change $u^p$ to achieve plausibility?  Assume $U_p$ is a metric space with metric $\rho_{U^p}$.  We can similarly define $\PR_{U^p}$ and $\FFM_{U^p}$ in terms of balls in $U^p$ of radius $d$ centered at $u^p$.  Plausibility robustness $\PR_{U^p}(u^p)$ requires that $\hb^r(\fb(u)) \in P$ for every $u \in B_{U^p}(u^p,d) \subseteq U^p$, and forced-fusion magnitude $\FFM_{U^p}(u^p)$ requires that $\hb^r(\fb(u)) \not \in P$ for every $u \in B_{U^p}(u^p,d) \subseteq U^p$.  Note that if $U^p = X^p$ and $\fb$ is the identity function, then these concepts can also be expressed directly in terms of $X^p$.

Now consider stating the producer's goal so that one might select $u^p \in U^p$ systematically, and even autonomously.  Perhaps the goal is to achieve a particular observation, say $y_G \in Y^r$, or more generally, any one in a nonempty set $Y_G$ of observations.  The producer must select $u^p$ so that $\hb^r(\fb(u^p)) \in Y^p$.  Further conditions might be that plausibility or illusions must be maintained by considering the resulting I-states (preimages of $h^r$).  Thus, the goal could alternatively be expressed directly in terms of I-states, which correspond to targeted perceptions: achieving an I-state $\is_G \in \ifs^r$ or set $\ifs_G \subset \ifs^r$ of I-states.



If there are multiple producer actions $u^p$ that achieve the goal, then an optimization problem can be formulated.  Let $l(u^p)$ be the cost of applying $u^p$.  An {\em optimal perception} (or equivalently, {\em optimal I-state}) would be the one that yields the minimum cost $l(u^p)$ over all $u^p$ that achieve the goal ($y_G$, $Y_G$, $\is_G$, or $\ifs_G$).  This assumes appropriate conditions for the existence of optima, such as compactness over the set of producer choices that achieve the goal.  Robustness and forced-fusion can also be taken into account.  For example, if $y^r \in P$, then $l(y^r) = c - \PR_{Y^r}(y^r)$, and if $y^r \not \in P$, then $l_{Y^r}(y^r) = c + \FFM(y^r)$, for some constant $c$.  An action can be chosen so that $l(y^r)$ is minimized among goal perceptions to maximize robustness; otherwise, it is chosen to minimize the forced-fusion magnitude.

The models extend naturally to handle disturbances.  In the nondeterministic case, $H^r \c X^r \ra Y^r$ and $\Hb^r \c \Omega \ra \pow(Y^r)$ are used instead of $h^r$ and $\hb^r$.  A disturbance could even be added to the producer's action so that $\fb$ is replaced by a function $\Fb \c U^p \ra \pow(X^p)$.  Thus, with each action, $u^p$, only a set $\Hb^r(\Fb(u^p)) \subseteq Y^r$ of possible observations can be enforced.  The producer is guaranteed to be successful in the worst-case if $\Hb^r(\Fb(u^p)) \subseteq Y_G$.  If $\Hb^r(\Fb(u^p)) \cap Y_G \ne \emptyset$, then it may {\em possibly} be successful, in the best-case.  Similarly, the case of probabilistic disturbance can be considered, using $p(y^r\mid x^r)$ and $p(y^r \mid \omega)$.  Disturbance can be added to producer actions as a pdf, $p(x^p \;|\; u^p)$.  The probability that a goal $Y_G$ is achieved is given by
\begin{equation}
p(Y_G\;|\;u^p) = \int_{Y_G} \int_{X^p} p(y^r\;|\;x^p) \; p(x^p \;|\; u^p) \; dx^p \; dy_G.
\end{equation}
Thus, the producer could try to choose $u^p \in U^p$ to maximize the probability of success.  For the disturbance-based models, costs could still be formulated.  In the nondeterministic case, $u^p$ can be chosen to minimize the worst-case (maximum) cost.  It the probabilistic case, $u^p$ can be chosen to minimize the expected cost.  The costs could include $\PR^r(y^r)$ and/or $\FFM(y^r)$, resulting in worst-case or expected-case analysis of plausibility robustness and forced-fusion magnitude.


\subsection{Creating Perceptual Experiences with an Omniscient Producer}\label{sec:op}


We now extend the concepts from Section \ref{sec:instant} from the stationary case to the dynamic case.  The extension of a stationary percept (or I-state) to cover multiple stages is called a {\em perceptual experience} (or I-state trajectory).  The producer and receiver are modeled using their corresponding 8-tuples, $\A^p$ and $\A^r$.  As before, the producer remains omniscient with $X^p = \Omega$, and $C^p$ modeling perfect correspondence.  It has access to the UTF $\fb \c \Omega \times U^p \times U^r \ra \Omega$, $\hb^r$, and $h^p = \hb^p$ is the identity function (perfect sensing).  It has access to $\is^r_k$, $u^r_k$, and $y^r_k$ at each $k \in \K$.  It also has access to the relations $C^r$ and $M^r$ (from which $R^r$ can be derived).  Relaxing these strong assumptions will be discussed in Section \ref{sec:imp}.

Consider going from stage $k$ to $k+1$.  The universe state is some $\omega_k \in \Omega$.  The producer can implement a state-feedback policy of the form $\pi^p \c \Omega \ra U^p$ (using the fact that $X^p = \Omega$).  Thus, an action $u^p_k = \pi^p(\omega)$ is selected.  The receiver action is $u^r_k = \pi^r(\is^r)$, and $\omega_{k+1} = \fb(\omega_k, u^p_k, u^r_k)$.  The next receiver observation is $y^r_{k+1} = \hb^r(\omega_{k+1})$.
The receiver I-state is updated as $\is^r_{k+1} = \phi^r(\is^r_k,u^r_k,y^r_{k+1})$, potentially using its intrinsic models $X^r$, $f^r$, and $h^r$.  Disturbance-based models could alternatively be used, including $F^r$ and $H^r$ or their probabilistic counterparts.

For every stage $k$, the producer applies $u^p_k$ to influence $y^r_{k+1}$ and $\is^r_{k+1}$.  This part is quite similar to Section \ref{sec:instant}, in which $u^p$ was chosen to influence $y^r$ and $\is^r$; however, here there is a one-stage delay.\footnote{Note that in the first stage $k=1$, the producer does not have the ability to affect $y^r_1$; this may be fixed by deleting the first observation or allowing the producer to start one stage earlier.}  Using $M^r$, the producer (or the engineer who created it) can determine whether each $\is^r_k$ is plausible.  Let $\histi^r$ denote an I-state sequence, called a {\em perceptual experience}, that is indexed over $k \in \K$.  If every $\is^r_k$ in $\histi^r$ is plausible, then it is called a {\em plausible perceptual experience}.  Let $\histo$ be the universe state trajectory corresponding to some $\histi^r$.  We can apply $M^r$ to determine whether each $(\is^r_k,\omega_k)$ is an illusion.  If the pair is an illusion for every $k \in \K$, then the pair $(\histi^r,\histo)$ is called an {\em illusory perceptual experience}.

\begin{example}[A Dynamic Landmark]\label{ex:1d2}
Building upon Example \ref{ex:1d}, let $\omega_k = (x^r_k,x^p_k)$ and $\omega_{k+1} = \fb(\omega_k,u^p_k,u^r_k) = (x^r_k + u^r_k, x^p_k+u^p_k)$, in which $U^p = U^r = \{-1,0,1\}$.  The landmark position can be moved up or down by one unit, or remain stationary, and the receiver can similarly change its own position.  We have $f^r(x^r_k,u^r_k) = x^r_k + u^r_k$ and $f^p(x^p_k,u^p_k) = x^p_k + u^p_k$.  Suppose $h^r$ functions as in Example \ref{ex:1d}.  Suppose $u^p_1 = 1$ and $u^r_1 = 0$.  This results in an implausible I-state $\is^r_2 = \emptyset$ because the position predicted by $f^r$ is inconsistent with the observation $y^r_2$ (the sets given by (\ref{eqn:ndet1}) and the preimage $(h^r)^{-1}(y^r_2)$ are disjoint).  To give the producer more freedom, nondeterministic replacements $F^r$ and $H^r$ can be used for the receiver's intrinsic models.  For example, if $F(x^r_k,u^r_k) = X^r$ for all actions, then the I-states would be based on preimages of $h^r$ alone, causing the receiver to have the illusion it is moving when in fact the producer is moving.  \qex


\end{example}

\begin{example}[Gridbot Illusions]\label{ex:gridboti}
Example \ref{ex:gridbot2} can be adapted by interpreting the gridbots as a producer and receiver.  The producer could create an illusion that the receiver's environment is smaller than it really is.  For example, suppose $E$ corresponds to two large rooms, connected by a `doorway' of width one tile.  The producer moves to the doorway and remains there while the receiver explores.  The illusion of a smaller room has been created.  If the producer moves away and the receiver returns to the doorway, an implausible I-state would result because a tile that was marked as black became white.  The challenge is to determine a policy for the producer so that it is not detected by the receiver; this would happen if the receiver sensed a tile as white when it was previously declared black, resulting in implausibility (assuming in its model it is unaware of the other robot).
\qex
\end{example}

Now consider designing a producer policy.  Under the standard agent model from Section \ref{sec:defs}, note that the entire system is predictable starting from any initial $\omega_1 \in \Omega$, which is immediately observable to the producer.  Call this the {\em fully predictable} case.   The producer policy may as well be a sequence of actions, which is a stage-feedback policy $\pi^p \c \K \ra U^p$, as defined in Section \ref{sec:ispaces}.  The goals from Section \ref{sec:instant} can be extended here to sequences.  For example, let $\histy_G \c \K \rightarrow Y^r$ be a {\em goal sequence} of receiver observations. A weaker requirement is to simply achieve any one $\histy_G$ in a set $\histY_G$ of possibilities.  Many other possibilities exist.  For example, perhaps the goal is to produce $y_G$, or any observation in a set $Y_G$, at {\em any} stage $k \in \K$.  Alternatively, perhaps it must happen at one particular stage.   A logic, such as linear temporal logic (LTL), may even be used to express goal conditions in terms of some combinations of sets of observations and stages \cite{FaiGirKrePap09}.  Furthermore, goals could also be expressed in terms of receiver states, receiver actions, receiver I-states, or any combination along with receiver observations and stages.  On top of this, a cost function $l_k$ can be defined at each stage to obtain a problem of finding an optimal producer action sequence, which in turn yields an {\em optimal perceptual experience} by optimizing the cumulative cost
\begin{equation}
\sum_{k \in \K} l_k(\omega_k, u^p_k, y^r_k, \is^r_k, u^r_k) .
\end{equation}
The costs can also be expressed in terms of $\PR$ and $\FFM$ functions to obtain problems that try to maximize total robustness or minimize total forced-fusion magnitude.  If $\K$ is infinite, then the costs must be carefully chosen so that the sum is finite for successful policies; alternatives include discounted cost, average cost, and termination actions \cite{Lav06}.  If $\K$ is finite, then a final cost term $l_F(\omega_{K+1},y^r_{K+1},\is^r_{K+1})$ may be added.

Now suppose that disturbance-based extensions of $f^r$ and $h^r$ are introduced for the receiver, to obtain $F^r$ and $H^r$, as defined in Section \ref{sec:defs}.  In this case, the receiver is no longer predictable, even from the producer's perspective.  It is thus more effective for the producer policy to be formulated as state-feedback $\pi^p \colon X^p \rightarrow U^p$, which even implies universe-state feedback.  Acknowledging that this may be extreme in many settings, Section \ref{sec:imp} removes producer omniscience to obtain other cases of information-feedback policies for the producer.

Consider characterizing the evolution of the whole system under the implementation of a fixed, state-feedback producer policy $\pi^p$.  Under the fully predictable case, a sequence $\histo \c \K \ra \Omega$ is determined from the initial universe state $\omega_0$.  In the case of a nondeterministic disturbance-based receiver, then a set $\histO$ of possible sequences is instead obtained.  If $\Omega$ is finite, then the process can be imagined as a nondeterministic finite automaton (NFA) over $\Omega$.  One should consider worst-case analysis to determine whether a goal can be guaranteed to be accomplished.  With a cost model, one can consider minimizing the worst-case perceptual experience.  In the case of a probabilistic disturbance-based receiver, a Markov chain is obtained under the implementation of $\pi^p$ (also called Markov decision process (MDP) by artificial intelligence researchers).  In this case, expected-case analysis could be used to assess the probability that the goal will be satisfied under $\pi^p$.  In this case, $\pi^p$ can be selected to maximize this probability. A cost model can additionally be used, with the resulting optimization being to find the lowest expected-case cost under the implementation of $\pi^p$.


\subsection{Producers with Imperfect Information}\label{sec:imp}

If the producer is not omniscient, then it may not have access to enough information to ensure that the targeted perceptual experiences function as desired.  To analyze what might happen between the producer and receiver, it will be helpful to nevertheless introduce a third-person perspective in which we as scientists or engineers have access to more information than the producer.  This could be modeled formally as an observer agent.

The following producer limitations can be considered: 1)  $\alpha^p$ and/or $h^r$ could be many-to-one mappings, prohibiting the producer from perfectly determining $\omega_k$;  2) $\pi^r$ may be unobservable to the producer, resulting in a {\em dynamic game} formulation, which may or may not be cooperative;  3) $\is^r$ may be partly or fully hidden, requiring the producer to estimate it via models, simulation, and limited sensors;  4) the producer may have only limited models or access to $h^r$, $\hb^r$, $f^r$, and $\fb$, resulting in partial control over illusions, and difficulty determining whether illusions are plausible.  5)  the producer may not have perfect access to its own state, leading to information feedback policies $\pi^p \c \ifs^p \ra U^p$ that hopefully achieve the targeted perceptual experiences.

Goals for targeted perceptual experiences may be formulated as in Section \ref{sec:op}, and a producer policy $\pi^p$ is selected that achieves the goal, and even optimizes costs.  For nondeterministic models, worst-case or even best-case analyses are appropriate.  For probabilistic models, expected-case analysis can be used once again.



\subsection{Multiagent Perceptual Experiences}\label{sec:multipe}

We can extend the formulations developed so far to allow for multiple producers and receivers within a shared universe.  Suppose, for example, that there is one producer and $n$ receivers.  The producer could use the same spoofing function $s$ to stimulate all of them at once.  This could be imagined as a broadcasting mechanism, such as wireless communication, that reaches all receivers in the same way.  Thus, the delivery of the perceptual experience could be considered as a kind of {\em centralized control}, applied by the producer.  We can alternatively formulate a {\em distributed control} scenario in which one or more producers deliver perceptual experiences and illusions propagate to receivers in a communication network.
One further extension is to allow multiple agents to be situated within a single body.  This could, for example, be used to model hierarchy, in which one agent plays a supervisory role over agents that function as lower-level models.  


\section{APPLICATION TO ROBOTS AND OTHER ENGINEERED AGENTS}\label{sec:robots}

\subsection{Modeling Engineered Agents}\label{sec:mea}


In engineering, we typically have white-box systems, which are built from well-understood physical principles and work as designed (as opposed to black-box systems, for which there is no understanding about their internals).  Learning, identification, and calibration processes may serve to further refine and improve their models with respect to their environment.  At a high level, any engineered agent can be modeled as a control system, including regulators of physical systems such as aircraft stability, room temperature, or the concentration of chemical solutions.  To cover many cases, a {\em linear agent} can be expressed as a discrete-time linear control system, which  can be formulated as $X = \Re^n$, $U = \Re^m$, XTE $x_{k+1} = A x_k + B u_k$, and sensor mapping $y_k = C x_k$ for fixed matrices $A$, $B$, and $C$.  State-feedback or information-feedback policies take the form $\pi \c X \ra U$ or $\pi \c \ifs \ra U$, respectively.  

Robot models typically involve nonlinear dynamical systems over configuration manifolds, often with non-trivial topology and non-Euclidean geometry.  Imagine extending the gridbot from Example \ref{ex:gridbot} to operate as a wheeled mobile robot, such as a robotic vacuum cleaner.  The configuration space $\C$ is used in robotics to model the set of all ways to embed the robot body in its environment in the absence of obstacles.  For a wheeled mobile robot, $\C$ could be the set of all 2D rigid body transforms: $\C = SE(2) \simeq \Re^2 \times S^1$, in which $S^1$ is the circle of all directions from $0$ to $2\pi$ (compare to $\Z^2 \times D$ from Example \ref{ex:gridbot}).  If the obstacles are unknown, then an environment space $\E$ could represent a set of possible subsets of $\C$ that are collision free, and then $X \subset \C \times \E$, in which $q \in \C$, $q \in E$, $E \in \E$ for any $(q,E) \in X$.  Actions $u_k \in U$ correspond to commands that cause the wheels to rotate for some time $\Delta t$, thereby altering the configuration  to obtain $x_{k+1} = f(x_k,u_k)$.

Onboard sensors are modeled by $h$ and might report whether obstacle contact is made, wheel odometry, and even distances to obstacles.  An onboard computer calculates I-states based on sensor observations.  The calculated I-states are used to apply actions according to a policy $u_k = \pi(\is_k)$.  Other types of robots, such as 3D drones, industrial manipulator arms, humanoids, or submarines, are similarly modeled using configuration spaces, $f$, and $h$.  To handle higher order dynamics, the configuration space may be extended to a higher-dimensional phase space that includes configuration velocities, and $f$ and $h$ are defined over it.

The result is a fully modeled, white-box system, which will highly contrast the modeling challenges for humans, discussed in Section \ref{sec:modhum}.  Nevertheless, in some settings robot models may be adaptive as action-observation pairs are accumulated during execution.  Models can be adjusted via machine learning or improved calibration.  A black-box setting may even appear if the engineer approaches an unknown robot, in which case perception engineering may be used to help understand its behavior.


\subsection{Spoofing Sensors}\label{sec:spoof}

Consider fooling sensors that might be used in robots.  The physical operation of $h$ in terms of the universe $\Omega$ is modeled as $\hb \c \Omega \ra Y$ (from Section \ref{sec:multi}), which allows a producer to create illusions for the receiver robot and result in $\hb^r(\omega) \ne h^r(\omx^r(\omega))$.  
Furthermore, obtaining plausible illusions and experiences for a robot is generally challenging because there are multiple sensors giving observations at multiple times.  All of these must be consistent with respect to a sensor fusion system in the sense that a possible receiver X-state trajectory could explain it in terms of $h$ and $f$.

{\em Localization} is the classic problem of estimating the robot's configuration.  If obtained by the trilateration system of Example \ref{ex:tri}, then localization illusions could be obtained by changing tower intensities.  Alternatively, the towers themselves could be moved.  Other wireless localization systems, including GPS, could be similarly spoofed.  {\em Mapping} is the problem of determining the robot's environment, usually in terms of obstacles that must be avoided; it is usually combined with localization to obtain {\em SLAM} \cite{ThrBurFox05}.  Many depth measurement systems work by emitter-detector pairs, such as sonars that emit a pulse and use the time of arrival to calculate distance.  Such sensors can be spoofed by blocking the emitted pulse, and sending an alternative pulse to the receiver at the desired time.  Methods for spoofing lidars for autonomous driving appear in \cite{ShiKimKwoKim17}.  Cameras that infer distance based stereo could be intentionally misaligned to give false results.  Features in images, assumed to be fixed in the world, could also be moved by a producer.  In an extreme case, a graphical display could even be placed in front of a camera.  Even a mechanical contact sensor, which is triggered when a robot hits a wall, could simply by fooled by pressing on it \cite{SuoNilLav20}.  

Mechanical sensors embedded in the body are the hardest to spoof.  For example, a modern inertial measurement unit (IMU) uses vibrating MEMS to estimate angular velocity and linear acceleration; this can be spoofed by injecting acoustic vibrations \cite{TriWeiXuHonFu17}.  {\em Odometry} and {\em joint encoders} report how far wheels have rolled and joints have rotated, respectively.  These are similar to proprioceptive senses in humans.  These could be spoofed by mechanical intervention, such as placing a mobile robot up on rack while the wheels rotate in the air.  This setup would use $\fb$ to additionally compensate for applied receiver actions so that plausible X-state transitions are nevertheless obtained.



\subsection{Virtual Reality for Robots}\label{sec:vrr}


\begin{figure}
\centerline{\hspace*{0.0in}\psfig{figure=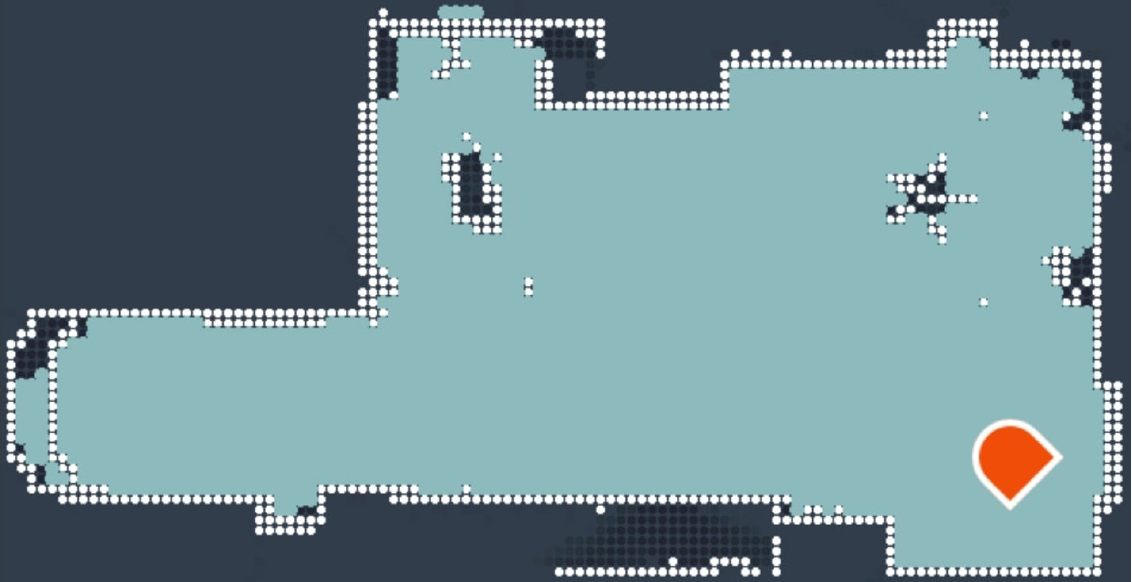,height=0.9in}\hspace*{-3.2in}\psfig{figure=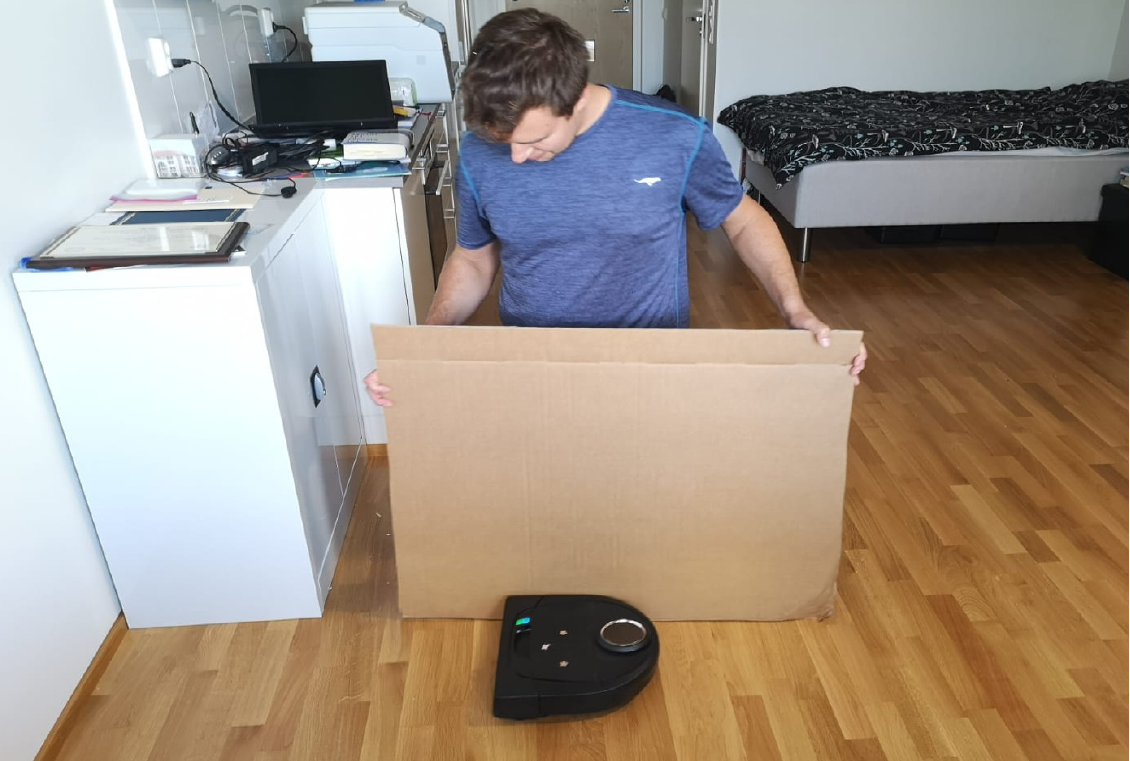,height=0.9in}\hspace*{-3.5in}\psfig{figure=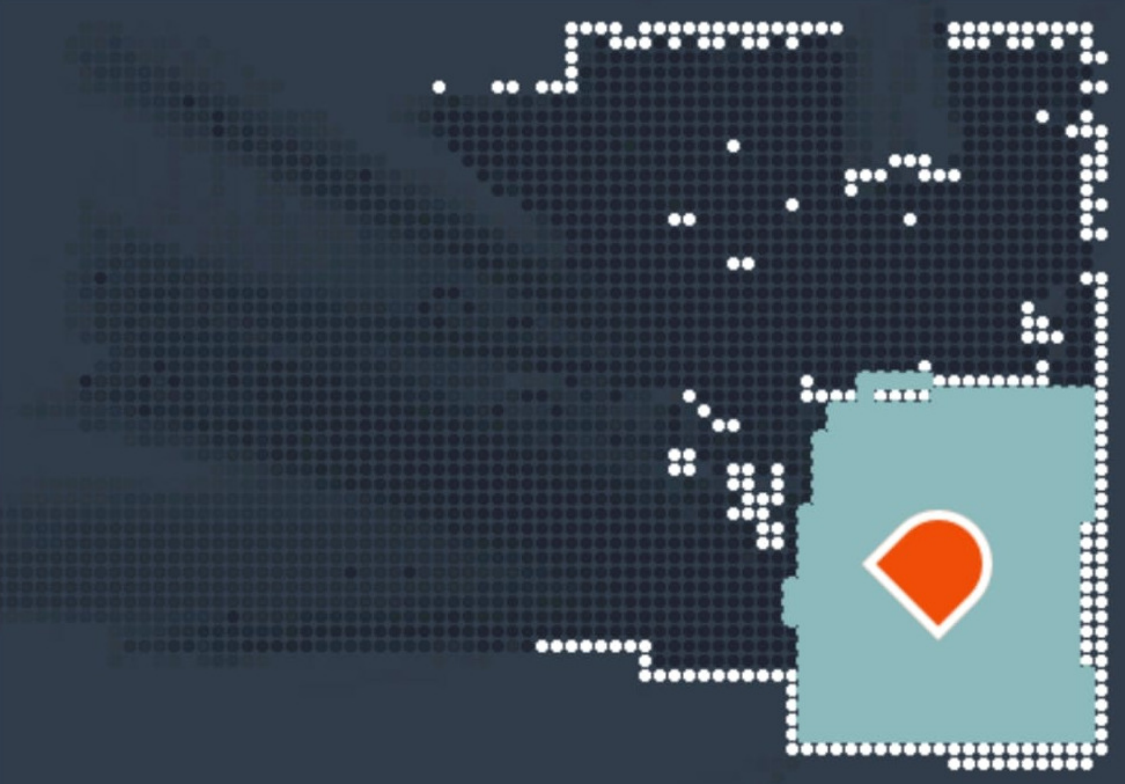,height=0.9in}}
\centerline{(a) \hspace*{1.5in} (b) \hspace*{1.3in} (c)}
\caption{\label{fig:mobrob} (a) A room mapped by a Neato Botvac D5 before interference.  (b) A producer (human) with cardboard causes sensor observations corresponding to a virtual wall.  (c) The receiver robot reports that it is done cleaning a smaller room than exists in reality (but its depth sensor measures some further away walls).
}
\end{figure}

With the ability to spoof sensors, we can next consider offering VR to an engineered agent, analogous to VR experienced by humans.  As a thought experiment, imagine a humanoid robot wearing a VR HMD.  Assuming it has cameras for eyes, the HMD might fool the sensor fusion system, though it is unlikely to work the same way as intended for humans.  The humanoid might walk and build an environment map that is consistent with the HMD imagery, with implausibility arising when it hits a real or virtual wall (the same would happen for a human using VR).  For a very different scenario, imagine offering VR, or a targeted perceptual experience, to a vacuum cleaning mobile robot; see Figure \ref{fig:mobrob}(a).  In \cite{SuoNilLav20}, researchers fooled the robot by moving a piece of cardboard around quickly so that the robot's wall contact sensor was activated as desired to create the illusion (Figure \ref{fig:mobrob}(b)).  The robot concluded it was in a smaller room than in reality, and it reported it was done cleaning (Figure \ref{fig:mobrob}(c)).  The mathematical framework of Section \ref{sec:core} covers these scenarios and many others in a unified way, although it remains to incorporate notions of one robot simulating another to create illusions \cite{SheOka20}.

Using concepts from Section \ref{sec:op}, suppose that some humans act as an omniscient producer.  The goal could be to get the robot into an I-state $\is^r$ in which it reports that its tasks have been accomplished.  This is achieved stage-by-stage by altering the physical world state $x^p$ so that that targeted observations $y^r$ are achieved for all sensors.  The particular $x^p$ chosen at each stage can depend on the sensed configuration of the receiver and even its I-states, if available.  Note that if it not possible to spoof all of the sensors, then the challenge of maintaining plausibility increases.

How are the targeted receiver observations determined?  One convenient way is to create a simulated world, which is then used to calculate what stimulus to provide to the sensors at each stage.  This is a way of maintaining a coherent, plausible `virtual' environment that responds to the receiver's actions and provides appropriate sensor feedback.  This is called a {\em virtual world generator} or {\em VWG}, and is a crucial component for VR applied to humans and other biological organisms \cite{Lav23}.  The simulator becomes a useful tool for maximizing plausibility robustness, or even determining whether plausibility is even possible.  If not, it might attempt to minimize the forced-fusion magnitude.  The computational complexity and ability of the simulator to respond in real time (within each stage) are important concerns.

In a robotics setting, we can even connect the 'brain' of the receiver directly to the simulator to evaluate planning, control, or sensor-fusion algorithms, as is done in software platforms such as Gazebo and CARLA.  The real-time requirement can even be neglected.  This would be the robot equivalent of a Gilbert Harman's `brain in a vat' (see \cite{Lav23}); however, VR for robots enables using the actual robot sensors in a physical setting to provide a more accurate assessment.  This could robot more thorough and systematic testing or verification of robot systems, and improve learning of models.  It also becomes possible to {\em reverse engineer} robots systematically by observing how they respond to various virtual environments.  This is quite analogous to the way neuroscientists and psychologists learn about the inner workings of biological organisms, to be discussed in Section \ref{sec:modhum}.

For a more challenging scenario, suppose the producer itself is a robot, which is constrained by its own $f^p$ and $h^p$.  It might need to do motion planning to each an appropriate $x^p$ to create the targeted observation $y^r$ at the precise stage it is needed.  It might have to choose trajectories that avoid detection by the receiver.  Imagine the challenges of getting a producer mobile robot to trick a vacuum cleaning receiver as was done in Figure \ref{fig:mobrob}, attempting to hit its wall contact sensors in the right places at the right time!  This results in a number of computational challenges, including deciding whether plausibility is possible, maximizing plausibility robustness, minimizing the forced-fusion magnitude, and determining the computational complexity of these problems in various settings.

\section{APPLICATION TO HUMANS AND OTHER BIOLOGICAL AGENTS}\label{sec:humans}

\subsection{Modeling Human Agents}\label{sec:modhum}

Although modeling engineered agents is challenging enough, it is much harder to model humans for several reasons: 1) Robots and other engineered agents are designed and built, component by component, using well-tested principles of physics, whereas biological agents simply exist.  They start as black boxes that are reverse engineered by scientists, including psychologists, neuroscientists, and biologists.  Thus, there is much speculation and debate about `what is going on' inside of the agent. 2) Data regarding internal operation can be easily collected during execution for engineered agents, but for humans we are limited to questionnaires and external biosensors such as those used in electroencephalography (EEG).  3) A major challenge for modeling humans is to maintain {\em ecological validity} in an experiment, but unfortunately, their behavior may be altered due to knowledge of participating in a study and other unnatural aspects of the experiment. 4) A robot can be simply rebooted for easy repeatability and to observe its reactions to varying environmental conditions; however, a human starts life only once and retains memories of prior trials.  Thus, his I-state cannot be re-initialized.  5) Humans {\em adapt} at various levels and time scales.  For example, eyes adjust to various lighting levels and people become more effective at using a computer mouse with practice.  Some adaption can be built into engineered systems, but it can also be avoided wherever preferable.  6) Through {\em attention processes} humans have the ability to ignore many things while focusing on others, thereby complicating what seems to be known at a particular moment.

One implication of these differences is far less repeatability or determinism for the case of biological agents, at least from the experimenter's viewpoint.  This has resulted in a preference for probabilistic models, and less ability to formulate an underlying deterministic model.  By contrast, engineered systems are typically modeled with a nominal deterministic part, and a stochastic part accounts for leftover disturbances.

For the worst-case, black-box extreme, imagine studying an impenetrable, mysterious gadget that was left behind by aliens.  We would have only our knowledge of physics, chemistry, and so on, to poke and prod it, and observe the results.  For modeling humans, we at least have useful models for human sensing and actuation.  For sensing, a sensor mapping of the form $h \c X \ra Y$ might model the human sense organs to a high degree of accuracy based on decades of research in physiology and neuroscience.  In this case, $X$ should include the possible stimuli to be presented to the organ, and $Y$ could be the resulting electrical impulses.  Vision is the most sophisticated sense and is complicated by many factors such as eye movements, pupil adaptation, optical distortions, and photoreceptor properties such as density, mosaics, response times, wavelength sensitivities, and amplitude sensitivities.  Furthermore, substantial neural processing occurs on the path from the retina through amacrine, horizontal, bipolar, and ganglion cells to the optic nerve.  All of these complicate $h$, making it imperfect and more challenging to model than a digital camera.  Other senses bring their own unique challenges: hearing, touch, thermoception, proprioception, pain, smell, taste, and vestibular.  In the actuation direction, we seek an XTE $f$ that yields the next external state $x_{k+1}$ as a function of the current state $x_k$ and a motor command $u_k$.  This involves modeling human body kinematics and dynamics; it falls under the field of kinesiology and includes the characterization of motor skills and learning.  Disturbance-based alternatives, such as $p(y_k \mid x_k)$ and $p(x_{k+1} \mid x_k, u_k)$, might be preferable.

Next imagine trying to extract a useful model of the human's I-space $\I$ and ITF $\phi$.  In the case of the alien gadget, its external state space $X$ can be systemically altered while observations about any physical changes the gadget undergoes are made, including movement or emitting energy such as lights or sounds.  A major assumption is that enough trials can be made with sufficient repeatability of behavior from the gadget, at least statistically.  To instead model a human, suppose that we can leverage acceptable models of $h$ and $f$.  In this way, variations in $X$ over the stages can be converted using $h$ and $f$ into hypothesized observations $y$ and actions $u$.  Thus, the brain appears to receive a history I-state.   This is consistent with most models in neuroscience, including for example, Friston's {\em free energy principle (FEP)}, in which the external and internal states (called I-states here) are separated by so-called {\em Markov blankets} \cite{Fri06}.   

Whether under the FEP or other models, the human executes a policy $\pi \c \ifs \ra U$; however, the problem is to characterize $\ifs$.  Setting $\ifs = \ifshist$ would make a brain with perfect memory and ability to make its motor commands to contingent on distinct history I-states.  Following Section \ref{sec:ispaces}, it is far more likely that an information mapping reduces $\ifshist$ to a sufficient, derived I-space $\ifsder$.  For most models used in neuroscience and perceptual psychology, this would be $\ifsprob$ from Section \ref{sec:ispaces}, and is consistent with the Bayesian brain hypothesis \cite{Gre80}, predictive coding \cite{RaoBal99}, and the FEP.

For a robot, the `brain' is a computer, for which any part of its internal state can be easily monitored.  By contrast, the human brain has around 86 billion neurons, with hundreds of millions more outside of the brain.  The operation of each neuron through axons and dendrites is itself a complicated dynamical system.  Direct measurement of neural activity is impossible, except in limited cases of single-unit recordings.  Instead, scientists must resort to non-invasive measures such as EEG, magnetoencephalography (MEG), and functional magnetic resonance imaging (fMRI).

Fortunately, humans can also be asked questions.  Thus, a common approach to modeling is {\em psychophysics} \cite{Fec88}, which aims to understand and quantify the relation between the I-states and the external world of physical stimuli by interactive questioning. Different psychophysical procedures, involving different types of tasks and settings can be used to target a certain aspect of the human visual system or, in general, a sensory system \cite{KinPri16,Tre95}. The human subject is asked to provide responses to questions based on provided stimuli.  The most basic procedures, with yes/no response, are stimuli detection and discrimination. Discrimination is the ability of tell two stimuli apart, whereas for detection one of the two stimuli is the `null' or `neutral' stimulus (for example, average luminance when detecting contrast sensitivity).

As mentioned above, attention processes \cite{BecKas09} are a major complication in the modeling of humans using VR.  They clearly have the knowledge that they entered VR, but nevertheless respond as if it is real \cite{meehan2002physiological} or they are present \cite{skarbez2017survey,slater2022separate}.  Even Slater's definition of a place illusion requires that the person knows he is someone else \cite{slater2022separate}.  This suggests that transitions in the ITF might vary based on attention.  It is as if there are multiple agents or I-spaces within one, with transitions affected by attention, which can be modeled as part of a high-level policy.


Finally, note that VR itself is a useful methodology to improve models of humans because scientists can observe their responses to carefully controlled, interactive experiences that would be difficult or impossible to produce in normal environments \cite{JeuHilBerGehGra22}.

\subsection{From Classical Illusions to Virtual Reality}\label{sec:ivr}

As mentioned in Section \ref{sec:intro}, artists have been creating perceptual illusions for millennia through paintings and sculpture.  By leveraging technological developments, modern artists who develop illusory perceptual experiences include skilled magicians, photographers, cinematographers, graphic artists, video game designers, and VR developers.  The term `illusion' is used somewhat loosely in everyday life, whereas the reality relation $R$ from Section \ref{sec:illusion} gave a precise definition.  Thus, we can apply it to well-known illusions to clarify what kind of illusions they are, or whether they should even be considered as illusions.

Suppose that a drawing or picture is shown to a human subject, and we ask her what she perceives.  The question could be constrained in a number of ways, such as providing multiple choices or asking whether one feature seems larger than another.  This setup can be modeled using the stationary formulation from Section \ref{sec:instant}.  The producer X-state $x^p = \omega$ places the picture in front of the human.  The observation $y^r$ models what is sensed by the receiver.  The I-state $\is^r$ is simply the reported answer to the question, ideally corresponding to what is perceived.  Using a reality relation $R^r$, we can determine whether various pairs $(\is^r,\omega)$ are illusions.  

Consider a line drawing of a rabbit, and $\is^r$ is perceiving a rabbit.  It is an illusion if $R^r$ is defined so that $\omega$ must correspond to a real rabbit being presented.  If the drawing is intended to be a rabbit and $R^r$ is defined accordingly, then it would {\em not} be an illusion to perceive a rabbit.  We must also determine whether $\is^r$ corresponds to perceiving an actual rabbit or a drawing of a rabbit.  In some cases, an illusion can be defined in a way that does not depend on the producer's intended interpretation.  Recall the so-called `illusions' of Figure \ref{fig:ill}.  For Figure \ref{fig:ill}(a), most people perceive the upper line segment to be longer than the lower one.  As an illusion of a 3D scene, it would be a longer embedded object.  However, as a line drawing, we can objectively measure the lengths of the two segments and conclude that they are the same length.  Thus, perceiving the upper segment as longer is an illusion in a measurable sense.  Similarly, the A and B tiles in Figure \ref{fig:ill}(b) have identical RGB values (when viewed on a screen), but an illusion of tile A being darker persists.  Figure \ref{fig:ill}(c) is different in that there is no objective ground truth regarding colors.  Some people see black and blue whereas many others see white and gold \cite{LafHerCon15}.  The reality relation can be defined in various ways based on what most people would interpret from the picture or even the original dress itself, but disagreement with what most other people would say hardly seems to be an illusion.  Figure \ref{fig:ill}(d) is an example of {\em multistable perception}, in which for most people the I-state oscillates over time between a rabbit and a duck.  Neither interpretation seems to be an illusion in the sense meant in this paper, unless one considers the fact that both are illusions because there is no real duck or rabbit present.  Obviously, $R^r$ could be defined in various ways, and future questions remain about which definitions are most reasonable or usefully capture intuitions about what is meant by an illusion.

Multistable perception yields varying I-states over time, even though the stimulus is stationary.  Now consider varying the stimulus by showing a sequence of pictures.  Imagine gradually increasing the rate of pictures shown per second.  Even at a few frames per second, we begin to perceive motion.  This illusion is known as {\em stroboscopic apparent motion} (see \cite{Lav23}) and is the basis of video media.  Another motion illusion is the {\em phi phenomenon} \cite{Wer12}, in which blinking dots around a circle induce a sense of motion.  The reality relation for these examples can be defined so that only true motion in $\Omega$ corresponds to reality; thus, such perceived motions are considered an illusion. 

An obvious limitation of motion pictures is their lack of interactivity.  This is overcome by video games, which take input from the user in the form of controllers and provide output in terms of video and audio displays.  A kind of {\em virtual world} is usually maintained in the game, which could be considered illusory using the concepts above.  Going a step further, VR creates a closed loop in which the person engages with a virtual world using more natural interaction mechanisms.  Wearing an HMD allows her to move her eyes, head, and body while seeing and hearing what the virtual world provides, while easily forgetting that it is a display.  Controlling the virtual world through natural methods such as hand movements and speech further increase the `realism' of the perceptual experience.  A {\em virtual world generator} maintains a consistent model as designed by the producer, and tracking systems estimate body configurations, to generate plausible responses \cite{Lav23}.

\begin{figure}
\centerline{\psfig{figure=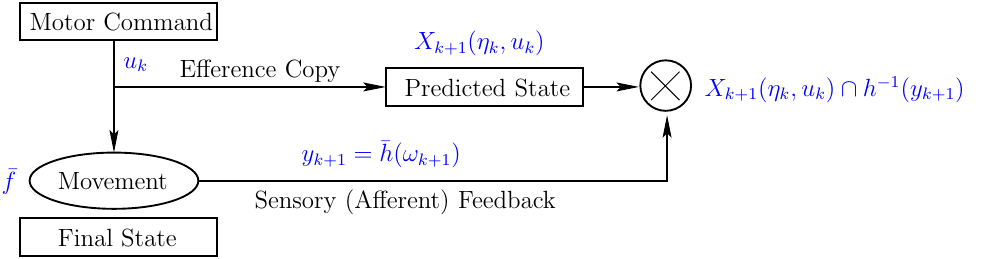,width=5.0in}}
\caption{\label{fig:smc} Sensorimotor contingency model from \cite{Gal00,GonLan17}, augmented with our nondeterministic I-state models. }
\end{figure}

A natural choice of explaining how VR works is the {\em sensorimotor contingency model} \cite{Gal00,OreNoe01,slater2022separate}, as proposed in \cite{GonLan17}.  Figure \ref{fig:smc} reproduces the model in \cite{GonLan17}, but adds our nondeterministic agent models from Section \ref{sec:ispaces}.  Suppose the receiver issues a motor command $u^r_k$.  The efference copy and I-state $X^r_k(\eta^r_k)$ used to calculate the `prediction' $X^r_k(\eta^r_k,u^r_k)$.  After movement, the next universe state is $\omega_{k+1} = \fb(\omega_k,u^r_k,u^p_k)$, and the sensor yields $y^r_{k+1} = \hb^r(\omega_{k+1})$.  If $X^r_k(\eta^r_k,u^r_k) \cap (h^r)^{-1}(y^r_{k+1}) \ne \emptyset$, then the I-state is plausible, as implied by the model relation $M^r$.  The reality relation $R^r$ could also applied to determine whether a perceptual experience is illusory.  This coarse model could be further detailed as multiple layers in a hierarchical predictive coding model, which has been generally successful in accounting for effects in visual perception \cite{Cla13,WalMcgClaOco20} and is theorized to explain the sensorimotor system as well \cite{BasUsrAdaManFriFri12,ShiAdaFri13}. 

A probabilistic formulation could also be made.  The prediction is $p(x^r_{k+1} \mid \eta^r_k, u^r_k)$, and instead of the preimage, the Bayesian posterior $p(x^r_{k+1} \mid y^r_{k+1})$ is calculated from a given $y_{k+1}$ and `uninformative' prior $p(x^r)$.  The degree of (im)plausibility could be expressed in terms of the {\em Kullback-Leibler (KL) divergence},
\begin{equation}\label{eqn:kld}
D_{KL}(p \;\|\; q) = - \int_X \log{(q(x)/p(x))} \; p(x) \; dx ,
\end{equation}
which represents an information-theoretic degree of `surprise'.  In (\ref{eqn:kld}), let $p = p(x^r_{k+1} \mid y^r_{k+1})$, which results from the observation, and $q = p(x^r_{k+1} \mid \eta^r_k, u^r_k)$ is the agent's prediction.  The model relation $M^r$ could define a binary-valued plausibility by setting a threshold on the KL divergence.  The KL divergence is also a critical component in the FEP.

In the design of VR systems, perception engineers would like to maintain or maximize plausibility while achieving a targeted perceptual experience.  The producer's `body' corresponds to the engineered artifacts that may be distributed throughout the environment, including displays, controllers, tracking systems, and so on.  Criteria to optimize include device expense, comfort, weight, development time, and adverse symptoms, such as fatigue or nausea.  The better each human sense is understood, the easier it is to build a VR display for it by exploiting its limits.  For example, a visual display need not be more than the maximum pixels-per-degree that are discernible by the human vision system.  Furthermore, a greater understanding of human perception and cognition leads to more opportunities to exploit their limitations in VR systems.  At an extreme, if $\ifshist$ is the brain model, then VR is as hard as possible because all histories lead to distinct perceptual experiences (I-state sequences).  Collapsing $\ifshist$ to smaller I-spaces enables more changes to be made to sensing and actuation that go unnoticed by the human.



\subsection{Evaluating Perceptual Experiences}\label{sec:epe}

Suppose that a user wears a VR HMD and has an engaging experience.  How can we measure whether the producer has successfully delivered a targeted perceptual experience?  What would be optimal, as discussed in \cite{skarbez2020immersion}?  This is difficult because the I-states are not directly accessible.  A detailed account of this process is provided in Chapter 12 of \cite{Lav23}.  As mentioned earlier, questionnaires can be designed to have people describe what they experienced.  Presence in a virtual environment is assessed using questionnaires, which are known to have unwanted biases \cite{Sla04,UsoCatArmSla00}.  Sickness symptoms have been assessed for decades using the {\em simulator sickness questionnaire} ({\em SSQ}) \cite{KenLanBerLil93}, which presents its own problems \cite{Law15}.  Physiological measurements can also be taken, but they are somewhat more cumbersome for the users because they must wear sensors, and the experience must be strong enough to yield a detectable response \cite{skarbez2017survey}.  



Sickness is one of the most challenging aspects to measure, and amounts to a set of symptoms including fatigue, nausea, dizziness, headache, and eyestrain \cite{Lav23,stanney2020identifying}.  It seems hard to find an analogous problem in robotic systems, except perhaps that an overheating CPU and power consumption may seem similar to fatigue.  Section \ref{sec:instant} introduced precise models of forced fusion (FFM) and plausibility robustness (PR), which could be used here to quantify the amount of mismatch or difficulty in maintaining plausibility.  This is related to sensory conflict theory \cite{oman1990motion}.  Even sickness symptoms in some cases can be modeled as a sum over stages of costs proportional to the FMM.  Mismatches that could be modeled mathematically within our framework include vergence-accommodation mismatch \cite{hoffman2008vergence}, visually induced motion sickness from vection \cite{KesHecLaw15}, flicker fusion \cite{ManMarWasShaKotWin21}, and even the mysterious {\em uncanny valley} phenomenon \cite{saygin2012thing}.  



It would be exciting if we could eventually develop new perceptual illusions that are directly predicted from perception engineering models.  Many clever techniques have been developed through understanding the limitations of human vision, such as foveated rendering \cite{levoy1990gaze}, frame skipping \cite{DenMarAshMan19}, and post-rendering image warp \cite{mark1997post}.  Redirected walking exploits limitations in human navigation to convince people they are walking straight when in fact they move in circles \cite{steinicke2009estimation}.  As examples continue to grow, how can approaches be more unified, with the systematic identification of many more?

\subsection{Social Perception Engineering}

Section \ref{sec:multipe} briefly described scenarios that involve multiple producers and/or receivers.  For humans, this leads naturally into {\em social} perception engineering.  At a very basic level, Shannon-Weaver communication can be modeled as a producer manipulating the environment (transmitting a message) that results in an intended I-state for the receiver (receiving the message).  From broadcasts to social media, information propagates among people.  Using reality relations for both producers and receivers, we can keep track of the spread of false information, a problem currently plaguing society as lies (a kind of illusion) are intentionally spread through a network of connected agents.  Note that only one producer is needed to create a lie, and the rest of the network may propagate it without awareness that it is an illusion.  Beyond the spread of messages, networked games (especially MMORPGs) and VR enable any number of people to interact through virtual worlds.  This leads to {\em transformed social interaction} \cite{BaiBeaLooBlaTur04}, in which people are able to have experiences that are different, and even better, than what could be accomplished in reality.  For example, imagine in an educational setting, a teacher can appear to be looking at every student at the same time.  People can design their own appearance or embodiment, so that biases based on physical characteristics such as race, gender, and height are readily overcome or studied by scientists under controlled conditions.




\subsection{Other Biological Organisms}\label{sec:others}


For organisms other than humans, VR is a rapidly maturing methodology for studying their behavior under controlled conditions \cite{naik2020animals}.  Since each organism has unique sensing and motor mechanisms, each VR system requires custom designs for displays and interaction methods.  As stated in \cite{naik2020animals}, there is an increasing need for engineers, especially roboticists and computer vision experts, to contribute to the development of VR systems for organisms, making VR for organisms an important part of perception engineering.  
Systems may be open-loop, such as showing visual stimuli to fish in an aquarium \cite{RosSovVal14}, or closed loop, as in \cite{ThuAya17}, in which a hamster runs on a ball while being presented with visual stimuli on a curved projection screen.  Scientists can learn about navigation, hunting, threat response, and many other behavioral aspects.  Even social behavior has been studied, for example in fish \cite{LarBai18,StoHofBasGri17}, to help understand swarming or schooling.  Their simpler neural structures and physiology often leads to models that have more detail and accuracy than is possible to obtain for humans.  VR-based methods for animals also provide better insights into brains, bodies, and behavior of humans.  Although questionnaires are not possible, more intrusive experimentation and measurement are allowed, such as conducting single-unit neural recordings.  It has even been established that place and grid cells, which are critical for human navigation, respond to VR experiences (for example, \cite{AghAchMooCusVuoMeh15}).  

Consider the challenges of constructing models of their I-spaces, and the particular I-states during the organisms VR-induced perceptual experience.  Starting from animals with the most human-like biology, studying macaque monkeys involves a VR setup that is unsurprisingly similar to that of humans \cite{SatSakTanTai04}, although single-unit recordings are at least possible.  For smaller animals, from rodents \cite{ThuAya17} to fruit flies \cite{StoHofBasGri17}, building a fully interactive virtual world is more feasible by having them run on rotating balls, or fly on a tether, and view fixed immersive displays (like miniature CAVE systems \cite{CruSanDefKenHar92}).  Single-unit recordings are performed in these settings to measure particular I-states.  Compared to 86 billion neurons for humans, fruit flies exhibit complex behaviors with only 100,000 neurons, resulting in a greater hope of fully understanding them.  Zebrafish larvae also have comparably many neurons and their bodies are transparent, allowing direct observation of their complete neural states \cite{KunLauMokKraKubFerForDal19}.  Although they have been studied in VR \cite{TriBol13}, it is more challenging to construct closed-loop systems in comparison to fruit flies and rodents.  VR has even been applied to roundworms \cite{FauRonThiLawMcCSotGriHecRobDoeLoc11}, for which its 302 neurons have been fully mapped \cite{cook2019whole}; however, measurement of I-states during execution remains challenging.  In \cite{Fea91}, a paramecium, which has no neurons, was manipulated into swimming along targeted trajectories by applying an electric field across the water; can the internal physical states of the paramecium be considered as I-states, resulting in a perceptual experience?  Perhaps not, and this example comes close to a point of debate among philosophers \cite{Fodor1986}.  Imagine a similar case of putting an object in a tray and using a robot to tilt the tray so that the object slides into a targeted configuration due to gravity \cite{ErdMas88}.  It would be ludicrous to claim that the object had a targeted perceptual experience, or even an illusion; nevertheless, the I-space concepts from Section \ref{sec:agents} are useful for designing `sensorless' manipulation strategies.

\section{CHALLENGES AND OPPORTUNITIES}\label{sec:end}

We have argued that perception engineering is an emerging discipline and introduced a mathematical framework to help characterize its scope and core.  In a general setting, we precisely characterized what it means for a producer to alter the environment to deliver an intended perceptual experience for a receiver, with the important conditions of it being plausible and illusory.  It will take decades of work to bring this envisioned discipline to maturity by a growing community of people to expand the foundations while leveraging important principles from other branches of engineering, and the sciences that study biological organisms.  This includes the need to train a generation of {\em perception engineers}.  If successful, perception engineering as a discipline would have a profound impact on society.  There are many exciting challenges and open problems ahead.  To conclude this article, we highlight some of these as a call to action.


The mathematical formulation of Sections \ref{sec:agents} and \ref{sec:core} should be expanded in several ways.  For example, they should account for more characteristics of biological organisms, especially adaptation and attention.  A continuous-time formulation using differential equations could be developed, which would naturally increase the connections to control theory.  One could define of optimal control problems, such as linear quadratic regulation (LQR) of perceptual experiences and illusions, in both discrete and continuous time.  One could adapt robust control techniques to targeted perceptual experiences.  Dynamic game theory can be applied to characterize equilibria between producers and receivers.  A control-theoretic model of psychophysics can be developed in which a feedback policy is used to guide actions that improve the agent modeling process.  Asynchronous and event-based models of sensing and interaction can be made, as opposed to having common stages.  The correspondence, model, and reality relations should be formulated in many more settings.  Logics should be applied to describe more complicated perception engineering tasks.  Finally, the universe space should be expanded to explicitly model how agents obtain information regarding I-states, actions, and observations.

Computational issues have barely been touched in this paper.  Under what conditions does a producer even exist that can achieve a targeted perceptual experience?  (This is analogous to computability in theoretical computer science.)  If it can, then what is the computational complexity of achieving it?  Can we find a minimal producer in some meaningful sense?  If the producer is a robot, then its mechanical capabilities and limitations must be taken into account.  If it is autonomous, then can planning methods be developed to achieve its goals, and with what complexity?  What forms of dynamic programming, including reinforcement learning, would be effective?  What can machine learning methods contribute to the development of better models, or what can machine learning gain from perception engineering?  What computational architectures and systems best support the creation of targeted perceptual experiences?  This itself is an emerging field of interest \cite{Huz22}.


Other fields of engineering are expected to benefit as well, including computer engineering and systems, computer graphics, sensing and vision systems, optical science, displays, acoustic engineering, filtering, and control theory.  Robotics should especially benefit because perception engineering can help to create better robots through improved modeling and learning techniques, and the development of robots that are more robust to spoofing attacks (attempts to create illusory perceptual experiences for robots).

The field of perception engineering should even contribute back to the sciences that study organisms, as is already the case for VR usage.  We expect to have improved understanding, definitions, and classifications of illusions.  We expect improved mathematical models of perception and cognition, which may be inspired by the considerable overlap between biological and engineered producers and receivers.  Finally, the mathematical formulations and engineering examples of perception engineering can help shed light on, and benefit from, continuing debates in the philosophy and cognitive science, especially involving situated agency, enactivism, semantics, symbol grounding, and 
representations \cite{Gallagher2017enactivist,Hiplito2022,HuttoMyin2012}.

\section*{DISCLOSURE STATEMENT}
The authors are not aware of any affiliations, memberships, funding, or financial holdings that
might be perceived as affecting the objectivity of this review.

\section*{ACKNOWLEDGMENTS}

This work was supported by the European Research Council (ERC AdG, ILLUSIVE: Foundations of Perception Engineering, 101020977), Academy of Finland (PERCEPT 322637, CHiMP 342556, PIXIE 331822), and Business Finland (HUMOR 3656/31/2019).

\bibliography{pe,vr}
\bibliographystyle{AR/ar-style3}


\end{document}